\documentclass[a4paper, 12pt]{article}
\usepackage{dsfont}

\usepackage{amsmath,amssymb,graphicx}
\usepackage{mathtools}
\usepackage{diagbox}
\usepackage{hyperref}
\usepackage[nosort]{cite}

\usepackage{tikz}
\usetikzlibrary{calc} \usetikzlibrary{patterns} \usetikzlibrary{decorations.pathreplacing} \usetikzlibrary{decorations.markings} \usetikzlibrary{decorations.pathmorphing} \usetikzlibrary{positioning}

\usepackage{color}
\definecolor{darkred}{rgb}{0.65,0.15,0}
\definecolor{newgreen}{rgb}{0.2,0.62,0.14}
\hypersetup{pdfborder={0 0 0},colorlinks=true,urlcolor=blue,citecolor=blue,linkcolor=darkred,linktocpage=true}

\DeclareFontFamily{U}{mathx}{}
\DeclareFontShape{U}{mathx}{m}{n}{ <-> mathx10 }{}
\DeclareSymbolFont{mathx}{U}{mathx}{m}{n}
\DeclareFontSubstitution{U}{mathx}{m}{n}

\DeclareMathAccent{\widecheck}{0}{mathx}{"71}

\numberwithin{equation}{section}

\newcommand{\starttext}{
\setcounter{footnote}{0}
\renewcommand{\thefootnote}{\arabic{footnote}}}

\def\nn{\nonumber}

\def\spa#1.#2{\left\langle#1\,#2\right\rangle}
\def\spb#1.#2{\left[#1\,#2\right]}

\def\half{{\scriptstyle \frac 12}}

\def\beq{\begin{equation}}
\def\eeq{\end{equation}}
\let\Re\relax

\DeclareMathOperator{\Re}{Re}

\def\cB{{B}}
\def\cC{{\cal C}}
\def\cG{{\cal C}}

\newcommand{\corr}{\mathcal{C}_N(\tau)}
\newcommand{\eisen}[1]{E^*\left(#1;\tau\right)}
\newcommand{\Dfn}[1]{D_N\left( #1;\tau\right)}
\newcommand{\Ymn}{Y_{mn}(\tau)}
\newcommand{\summn}{\sum_{(m,n)\neq (0,0)}}
\newcommand{\NPfn}[1]{\mathcal{E}(#1;\tau)}
\newcommand{\tFo}[4]{{}_2F_1\left(#1,#2;#3\vert #4\right)}
\newcommand{\tFt}[6]{{}_3F_2(#1,#2,#3;#4,#5\vert #6)}
\newcommand{\PhiP}[1]{\mathcal{C}^{(#1)}_P(N;\tau)}
\newcommand{\PhiNP}[1]{\mathcal{C}^{(#1)}_{N\!P}(N;\tau)}

\newcommand{\polylog}[2]{\text{Li}_{#1}(#2)}
\newcommand{\average}[1]{\langle #1\rangle}
\newcommand{\correl}[1]{\mathcal{C}_{#1}(\tau)}

\newcommand{\intmed}{\int_{\mathcal{M}}}

\newcommand{\eq}{\begin{equation}}
\newcommand{\eqe}{\end{equation}}
\newcommand{\eqa}{\begin{eqnarray}}
\newcommand{\eqae}{\end{eqnarray}}

\newcommand{\bea}{\begin{eqnarray}}
\newcommand{\eea}{\end{eqnarray}}
\newcommand{\dd}{\mathrm{d}}

\newcommand{\ZZ}{\mathbb Z}
\newcommand{\Z}{\mathbb Z}

\newbox\charbox
\newbox\slabox
\def\s#1{{      
        \setbox\charbox=\hbox{$#1$}
        \setbox\slabox=\hbox{$/$}
        \dimen\charbox=\ht\slabox
        \advance\dimen\charbox by -\dp\slabox
        \advance\dimen\charbox by -\ht\charbox
        \advance\dimen\charbox by \dp\charbox
        \divide\dimen\charbox by 2
        \raise-\dimen\charbox\hbox to \wd\charbox{\hss/\hss}
        \llap{$#1$}
}}

\usepackage[shadow,textwidth=2.7cm]{todonotes}
\usepackage{ifthen}
\reversemarginpar

\begin{document}

\starttext

\setcounter{footnote}{0}

\begin{flushright}
{\small }
\end{flushright}

\vskip 0.3in

\begin{center}

\centerline{\large \bf  Large-$N$ integrated correlators in $\mathcal{N}=4$ SYM: }
\vskip 0.1cm
\centerline{\large \bf  when resurgence meets modularity } 

\vskip 0.3in

{ Daniele Dorigoni and Rudolfs Treilis} 
   
\vskip 0.3in

{\small  \it{Centre for Particle Theory \& Department of Mathematical Sciences,} 
}\\
\small{\it{ Durham University, Lower Mountjoy, Stockton Road, Durham DH1 3LE, UK}}

\vskip 0.5in

\begin{abstract}
 
 \vskip 0.1in

Exact expressions for certain integrated correlators of four half-BPS operators in $\mathcal{N} = 4$ supersymmetric Yang--Mills theory with gauge group $SU(N)$ have been recently obtained thanks to a beautiful interplay between supersymmetric localisation and modular invariance. 
The large-$N$ expansion at fixed Yang--Mills coupling of such integrated correlators produces an asymptotic series of perturbative terms, holographically related to higher derivative interactions in the low energy expansion of the type IIB effective action, as well as exponentially suppressed corrections at large $N$, interpreted as contributions from coincident $(p, q)$-string world-sheet instantons.
In this work we define a manifestly modular invariant Borel resummation of the perturbative large-$N$ expansion of these integrated correlators, from which we extract the exact non-perturbative large-$N$ sectors via resurgence analysis. Furthermore, we show that in the 't Hooft limit such modular invariant non-perturbative completions reduce to known resurgent genus expansions.
Finally, we clarify how the same non-perturbative data is encoded in the decomposition of the integrated correlators based on $\rm{SL}(2,\mathbb{Z})$ spectral theory.

\end{abstract}                                            
\end{center}

\baselineskip=15pt
\setcounter{footnote}{0}

\newpage

\setcounter{page}{1}
\setcounter{tocdepth}{2}
\tableofcontents

\bigskip

\section{Introduction}
\label{sec:1}

Perhaps one of the most intriguing quantum field theories in four space-time dimensions is $\mathcal{N}=4$ supersymmetric Yang-Mills theory (SYM) with gauge group $SU(N)$. Amongst the many reasons for its undeniable appeal is that it provides for a moduli space of non-trivial superconformal theories parameterised by the Yang--Mills coupling $g_{_{Y\!M}}$ and theta angle $\theta$, conveniently packaged in the complex coupling $\tau\coloneqq {\theta}/(2\pi)+ {4\pi i}/{g_{_{Y\!M}}^2}$, upon which the Montonen--Olive \cite{Montonen}  duality group $\rm{SL}(2,\ZZ)$ acts.

Another important reason for our interest in $\mathcal{N}=4$ SYM is that it provides a non-perturbative description of type IIB string theory in an $AdS_5\times S^5$ background \cite{Maldacena:1997re}.  The gauge theory coupling $\tau$ is identified holographically with the type IIB string theory coupling $\tau_s\coloneqq \chi+i/g_s$, i.e.~$\tau =\tau_s$, while the string length scale $\ell_s$  is related to the number of colours $N$ of the dual $SU(N)$ gauge theory side by $(L/\ell_s)^4= g_{_{Y\!M}}^2 N$, with $L$ the length scale of the $AdS_5\times S^5$ background.

Thus, the string theory effective gravitational description at small $\ell_s$ corresponds on the gauge side to a large-$N$ and finite $\tau$ regime. Although in this limit the bulk is weakly curved, as a consequence of $\ell_s/L\to 0$, the string theory remains strongly coupled due to $\tau_s$ being finite. Super-graviton scattering amplitudes on $AdS_5\times S^5$ are consequently related to correlation functions of the stress tensor multiplet in $\mathcal{N}=4$ SYM. Despite offering a non-perturbative definition of string theory through a well-defined CFT, practical applications of this duality aimed at exploring string theory in the gravity regime remain challenging since the CFT is still strongly coupled at large $N$. To advance our understanding of quantum gravity through AdS/CFT it  is therefore necessary to analyse $\mathcal{N}=4$ SYM non-perturbatively.

An extremely powerful method to extract non-perturbative properties of $\mathcal{N}=4$ SYM for arbitrary coupling $\tau$ and number of colours $N$, is the use of supersymmetric localisation. 
It is precisely thanks to this tool that it was recently understood \cite{Binder:2019jwn,Chester:2020dja} how to obtain certain integrals of the correlator of four superconformal primary operators, usually denoted\footnote{For simplicity we suppress $R$-symmetry indices.} by $\mathcal{O}_2(x)$, in the $\mathcal{N}=4$ stress tensor multiplet.

The particular integrated correlators of interest for the present work have been introduced in  \cite{Binder:2019jwn,Chester:2020dja} and are computed from derivatives of the $S^4$ partition function $Z_N(m,\tau)$ for $\mathcal{N}=2^*$ SYM obtained by Pestun \cite{Pestun:2007rz} in terms of an $SU(N)$ matrix model integral:
\begin{equation}
\label{eq:GHdef}
\mathcal{C}_N(\tau)\coloneqq \frac{1}{4}\Delta_\tau \partial_m^2\log Z_N(m,\tau)\big\vert_{m=0}\,,\qquad \mathcal{H}_N(\tau)\coloneqq \partial_m^4\log Z_N(m,\tau)\big\vert_{m=0}\,,
\end{equation}
where $\Delta_\tau\coloneqq 4 \tau_2^2\partial_\tau\partial_{\bar\tau}$ is the hyperbolic laplacian and the complexified coupling constant $\tau=\tau_1 + i \tau_2\in \mathfrak{H}$ parametrises the upper-half plane $ \mathfrak{H} \coloneqq \{\tau\in \mathbb{C}\,:\,\mbox{Im}(\tau)>0\}$, where we identify $\tau_1={\theta}/{(2\pi)}$ and $\tau_2={4\pi}/{g_{_{YM}}^2}$. 

When the mass parameter $m$ is set to zero $\mathcal{N}=2^*$ SYM reduces to $\mathcal{N}=4$, hence the expressions just defined correspond to $\mathcal{N}=4$ observables.
More precisely, the quantities $\mathcal{C}_N(\tau)$ and $\mathcal{H}_N(\tau)$ are identified with integrals over the insertion points of four superconformal primary operators of the form,
\begin{equation}
\label{eq:GHInt}
\mathcal{C}_N(\tau)=\int \langle \mathcal{O}_2(x_1) \cdots \mathcal{O}_2(x_4)\rangle\,\dd \mu(\{x_i\})  \,,\qquad \mathcal{H}_N(\tau)=\int  \langle \mathcal{O}_2(x_1) \cdots \mathcal{O}_2(x_4)\rangle\,\dd \tilde{\mu}(\{x_i\})\,.
\end{equation}
We refer to~\cite{Binder:2019jwn,Chester:2020dja} for the precise relation between the supersymmetric localisation definitions \eqref{eq:GHdef} and the exact forms of the integrated correlators \eqref{eq:GHInt}, and in particular for details on the integration measures~$\dd \mu(\{x_i\})$ and~$\dd \tilde{\mu}(\{x_i\})$ distinguishing~$\mathcal{C}_N(\tau)$ and~$\mathcal{H}_N(\tau)$.

Thanks to these results, it has finally become possible to perform holographic ``precision-tests'' \cite{Binder:2019jwn,Chester:2019jas,Chester:2020dja,Chester:2020vyz,Alday:2021vfb}, and reconstruct from the large-$N$ expansion of the integrated correlators \eqref{eq:GHdef}, the first few low-energy string theory corrections to the tree-level supergravity contribution to four-graviton scattering in $AdS_5\times S^5$ as well as in flat-space.
 
Surprisingly, in a series of papers~\cite{Dorigoni:2021bvj,Dorigoni:2021guq, Dorigoni:2022zcr} an exact and modular covariant expression for finite $\tau$ was proven for a generalisation of the first integrated correlator $\mathcal{C}_N(\tau)$ to arbitrary classical gauge group, then extended to exceptional gauge groups in \cite{Dorigoni:2023ezg}. A key r\^ole in determining these astonishing results is played by the action of the Montonen--Olive duality group $\rm{SL}(2,\ZZ)$ on the complex coupling $\tau$, strongly constraining the space of modular invariant objects at play.

This led to a flourishing of exact results\footnote{See \cite{Dorigoni:2022iem} for a recent review.} for other integrated correlators in $\mathcal{N}=4$ SYM such as higher-point maximal $U(1)_Y$-violating correlators~\cite{Green:2020eyj, Dorigoni:2021rdo}, four-point functions of higher conformal dimensions operators \cite{Paul:2022piq,Brown:2023cpz,Paul:2023rka,Brown:2023why} and giant gravitons \cite{Brown:2024tru},  as well as integrated two-point functions of two superconformal primary operators in the presence of a half-BPS line defect~\cite{Pufu:2023vwo,Billo:2023ncz}. More recently these methods have also been applied to integrated correlators in less supersymmetric theories such as~$\mathcal{N}=2$  SYM~\cite{Billo:2023kak,Behan:2023fqq,Pini:2024uia}. We stress that these finite-$N$, finite-coupling results provide furthermore important data for numerical bootstrap studies, see e.g. \cite{Chester:2021aun,Behan:2024vwg}.

While novel studies \cite{Alday:2023pet} have shown intriguing conjectural relations between the two integrated correlators in \eqref{eq:GHdef}, particularly in the large-$N$ fixed-$\tau$ limit, we still lack an exact modular invariant expression for the second integrated correlator $\mathcal{H}_N(\tau)$ valid for arbitrary $N$ and fixed $\tau$. On the contrary, thanks to the pivotal results of \cite{Dorigoni:2021bvj,Dorigoni:2021guq} we have an almost complete control over the first integrated correlator $\mathcal{C}_N(\tau)$.
In particular we know that for all $N$ the quantity $\mathcal{C}_N(\tau)$ can be represented as a simple lattice sum integral whose systematic large-$N$ expansion can be computed \cite{Dorigoni:2022zcr} starting from a lattice sum generating series over the number of colours $N$. 

From the analysis of \cite{Dorigoni:2022zcr} it follows that the large-$N$, fixed-$\tau$ expansion of $\mathcal{C}_N(\tau)$ is an asymptotic factorially divergent formal series which has to be completed by an infinite tower of modular invariant, non-perturbative exponentially suppressed terms at large-$N$, thus confirming the earlier impressive numerical studies of \cite{Hatsuda:2022enx}. These non-perturbative corrections are extremely important and have the holographic interpretation of contributions from coincident $(p, q)$-string world-sheet instantons.

The first main scope of this paper is to provide a resurgence analysis approach to the resummation of modular invariant large-$N$ perturbative expansions akin to that for $\mathcal{C}_N(\tau)$. We show that it is possible to define a modified Borel resummation kernel with manifest modular invariance. By applying this resummation procedure to the formal perturbative large-$N$ expansion of the integrated correlator $\mathcal{C}_N(\tau)$, we retrieve its complete exact transseries expansion previously only found via generating series methods. 
The modular invariant non-perturbative sectors of $\mathcal{C}_N(\tau)$ are amazingly encoded in its perturbative part.
We also show that our approach is extremely useful in deriving novel non-perturbative results for a particular sector of the large-$N$ expansion for the second integrated correlator $\mathcal{H}_N(\tau)$.
The proposed modular invariant resurgent resummation is furthermore perfectly suited to perform a large-$N$ 't Hooft limit expansion at large $\lambda\coloneqq N g_{_{Y\!M}}^2$, thus recovering the non-perturbative worldsheet instanton completions first obtained using resurgence analysis order by order in the genus expansion \cite{Dorigoni:2021guq, Hatsuda:2022enx}. 

The second parallel, yet deeply intertwined goal of this work is understanding how the large-$N$ exact transseries expansion of the integrated correlator~$\mathcal{C}_N(\tau)$ is encoded in an alternative and equivalent representation nicely found in~\cite{Collier:2022emf} via~$\rm{SL}(2,\mathbb{Z})$ spectral theory. The key idea behind spectral analysis is that modular invariant functions,~such as~$\mathcal{C}_N(\tau)$, can be decomposed as linear combinations of “good” basis elements,~i.e.~$L^2$-normalisable eigenfunctions of the hyperbolic Laplace operator~$\Delta_\tau$.
We show that the large-$N$ expansion of this spectral decomposition yields precisely the spectral decomposition of the large-$N$ transseries expansion obtained via resurgence analysis.

The outline of the paper is as follows.
In section \ref{sec:review} we review some important properties of the integrated correlator $\corr$, in particular we present its lattice sum representation and generating series, thanks to which the large-$N$ modular invariant transseries expansion was first derived. We also present a spectral representation for $\corr$ obtained via $\rm{SL}(2,\mathbb{Z})$ spectral theory.
In section \ref{resurgence_section} we define a modular invariant resummation for formal large-$N$ perturbative expansions akin to that of $\corr$, and demonstrate how this method can be used to derive non-perturbative but modular invariant corrections at large-$N$.
We present how this resummation method can be used to reconstruct both the full transseries expansion of $\corr$ as well as a particular non-perturbative sector of the second integrated correlator $\mathcal{H}_N(\tau)$.
In the process, we show that our modular invariant resurgent resummation neatly encodes the large-$N$ 't Hooft limit transseries expansion of~$\corr$.
The same modular invariant large-$N$ transseries expansion of $\corr$ is derived in section \ref{spectral_section} from a spectral representation for the integrated correlator, thus clarifying how the large-$N$ non-perturbative sectors are encoded in the spectral overlap functions.
We conclude in section \ref{sec:Conc} with some comments on possible future directions. The paper contains two appendices where we discuss more technical details.

\section{Brief review of integrated correlators}
\label{sec:review}

In this section we review some of the key properties of the integrated correlator $\corr$ defined in \eqref{eq:GHdef} which will be the main character of our story.
In particular, we present two equivalent representations valid for all complex coupling $\tau\in \mathfrak{H}$ and arbitrary number of colours $N$. Firstly in section \ref{sec:Lattice}, we discuss the original exact expression for $\corr$ found in \cite{Dorigoni:2021bvj,Dorigoni:2021guq} in terms of a lattice sum combined with a Borel-like integral transform. In section \ref{sec:TS} we review how to construct from the lattice-sum representation a generating series over $N$ and with it compute the exact large-$N$ modular invariant transseries \cite{Dorigoni:2022cua}. 

Secondly, in section \ref{sec:spec} we discuss an equivalent expression for the integrated correlator $\corr$, {first} presented in \cite{Collier:2022emf} and then extended in \cite{Paul:2023rka}, in terms of an extremely simple spectral representation with respect to $L^2$-normalisable functions on the fundamental domain of $\rm{SL}(2,\mathbb{Z})$.
One of the main results of our work is showing how the spectral data beautifully encodes the large-$N$ modular invariant perturbative expansion of \eqref{eq:GHdef}, its non-perturbative completion and, in fact, the complete transseries representation.

\subsection{Lattice sum representation}
\label{sec:Lattice}

Even though the integrated correlator $\corr$ is defined in  \eqref{eq:GHdef} by taking a suitable combination of derivatives of the $S^4$ partition function in the $\mathcal{N}=2^*$ mass deformed supersymmetric Yang--Mills theory, it was proven in \cite{Dorigoni:2021bvj,Dorigoni:2021guq} that this integrated correlator has the far more convenient lattice sum representation,
\begin{equation}\label{corr_int_definition}
    \corr = \frac{1}{2}\sum_{(m,n)\in \mathbb{Z}^2}\int_0^\infty e^{-t\,\Ymn}  B_N(t)  \,\dd t\,,
    \end{equation}
    where we define the ubiquitous ``lattice-sum coupling'',
\begin{equation}
\Ymn \coloneqq \pi \frac{|n\tau+m|^2}{\tau_2}.
\end{equation}
This extremely simple formula can be seen as a combination of the lattice sum over $(m,n)\in \mathbb{Z}^2$ and a Laplace integral of a ``Borel transform'' function $B_N(t)$, which is a rational function of $t$ given by
 \begin{equation} \label{eq:BSUN}
 B_{N}(t)\coloneqq \frac{\mathcal{Q}_{N}(t)}{(t+1)^{2N+1}}\, .
 \end{equation} 
 The function  $\mathcal{Q}_{N}(t)$ is a polynomial in the variable $t$ of degree $2N-1$ which can be written for all $N\in \mathbb{N}$ as
\begin{align}
 \mathcal{Q}_{N}(t) &\label{eq:Bndef} \coloneqq -{\frac{1}{2}}N(N-1)(1-t)^{N-1}(1+t)^{N+1}\\
 &\nn \left\lbrace [3+(3t+8N-6)t] P_N^{(1,-2)}\Big(\frac{1+t^2}{1-t^2}\Big) +\frac{1}{1+t} (3t^2-8 N t-3 )P_{N}^{(1,-1)}\Big(\frac{1+t^2}{1-t^2}\Big)\right\rbrace\,,
 \end{align}
 where $P_n^{(a,b)}(x)$ are Jacobi polynomials. 
 We note that for any $N$ the function $B_N(t)$ satisfies the inversion identity 
\begin{equation}\label{B_inversion_identity}
    t^{-1}B_N(t^{-1})=B_N(t)\,.
\end{equation}
 The form of the function $B_{N}(t)$ given in \eqref{eq:BSUN}  and  \eqref{eq:Bndef} was conjectured  in \cite{Dorigoni:2021guq, Dorigoni:2022zcr} and then proved in \cite{Dorigoni:2022cua} using matrix model methods.
 
Interestingly, in \cite{Dorigoni:2022zcr} it was shown that a more general version of the lattice sum expression \eqref{corr_int_definition}, yields the integrated correlator of four superconformal primary operators \eqref{eq:GHInt} for $\mathcal{N}=4$ SYM with arbitrary classical gauge group $G=SO(N),USp(2N)$, then completed in \cite{Dorigoni:2023ezg} to the case of exceptional gauge groups. Goddard-Nuyts-Olive \cite{Goddard:1976qe} electro-magnetic duality plays a fundamental r\^ole in dictating the particular lattice sum expressions appearing for different gauge groups.
For the rest of this paper we  focus our attention to the original discussion \eqref{corr_int_definition} of the integrated correlator in the $SU(N)$ theory.

As already noted, compared to the original expression \eqref{eq:GHdef}, it is much easier to analyse the dependence of the integrated correlator from the parameters $\tau$ and $N$ starting from the lattice sum expression \eqref{corr_int_definition}. However, while the $\tau$ dependence has been basically trivialised, the dependence of \eqref{eq:BSUN} on the number of colours $N$ is absolutely not transparent.  
This shortcoming was remedied in \cite{Dorigoni:2022cua} where a generating function for the $N$-dependence was derived starting from \eqref{corr_int_definition}, thus allowing for a direct calculation of the exact large-$N$, fixed-$\tau$ transseries expansion.

This generating series is defined as
  \begin{equation}
 \label{eq:gendef}
 \cG_{SU}(z;\tau ) := \sum_{N=1}^\infty   \cC_{N}(\tau)\,z^N\,,
 \end{equation}
 with $z$ an auxiliary complex variable.
We can then invert \eqref{eq:gendef} via
\begin{equation}
\label{eq:geninv}  
 \cC_{N}(\tau)\  = \oint_\gamma \frac{\cG_{SU}(z;  \tau)}{z^{N+1}}  \frac{\dd z}{2\pi i } \,,
\end{equation}
 where $\gamma$ denotes a counter-clockwise contour circling the pole at $z=0$ of radius strictly less than one in order to avoid other singularities.
  From \eqref{corr_int_definition} we can equivalently define the generating series for  the rational functions $B_{N}(t)$
which can be computed directly from \eqref{eq:BSUN}:
 \begin{equation}
 \cB_{SU}(z;t) := \sum_{N=1}^\infty B_{SU(N)}(t) z^N = \frac{3 t z^2 \left[(t-3) (3 t-1)(t+1)^2 -
   z(t+3) (3 t+1) (t-1)^2 \right]}{(1-z)^{\frac{3}{2}}
   \left[(t+1)^2-(t-1)^2 z\right]^{\frac{7}{2}}} \, ,
   \label{eq:gresult}
  \end{equation}
leading to
  \begin{equation} \label{eq:cGSU}
  \cG_{SU}(z;\tau) := {\frac{1}{2}}\sum_{(m,n)\in\Z^2} \int_0^\infty e^{-t \, Y_{mn}(\tau)}  \cB_{SU}(z;t) \, \dd t\,.
  \end{equation}
  
 This generating function satisfies several properties of note,
 \begin{equation}
 \cB_{SU}(z;t) = t^{-1} \cB_{SU}(z; t^{-1})\,, \qquad\qquad 
 \cB_{SU}(z;t) =-  \cB_{SU}(z^{-1}; -t)  \,,\nn 
 \end{equation}
 as well as the integral identities,
 \begin{equation}
 \qquad\quad \int_0^\infty \frac{ \cB_{SU}(z;t)}{\sqrt{t}} \, \dd t = 0\,,\qquad\quad\quad
 \int_0^\infty  \cB_{SU}(z;t) \,\dd t 
 = \sum_{N=1}^\infty \frac{N(N-1)}{{4}} z^N\,.
 \label{eq:propGSU2}
 \end{equation}
 The first of these equations, directly related to \eqref{B_inversion_identity}, is an inversion relation that follows automatically from the lattice sum definition of the integrated correlator \eqref{corr_int_definition}, as was pointed out in \cite{Collier:2022emf} where the lattice sum is re-expressed in terms of a modular invariant spectral representation which will shortly be reviewed.
The second equation in \eqref{eq:propGSU2} is an inversion relation in the variable $z$, which relates the $SU(N)$ correlator with coupling $g_{_{Y\!M}}^2$ to the $SU(-N)$ correlator with coupling $-g_{_{Y\!M}}^2$, as previously discussed in \cite{Dorigoni:2022zcr}.

\subsection{Modular invariant large-$N$ transseries}
\label{sec:TS}

One of the main advantages of introducing a generating series such as $\cG_{SU}(z;  \tau)$ is that it has a much simpler form than $\cC_{N}(\tau)$. This makes $\cG_{SU}(z;  \tau)$ extremely convenient for analysing the large-$N$ properties of the integrated correlators.   
In particular, starting from \eqref{eq:cGSU} a key result of \cite{Dorigoni:2022cua} was the derivation of the exact large-$N$ transseries expansion for $ \cC_{N}(\tau)$ at fixed $\tau$, which takes the form
\begin{equation}
\cC_{N}(\tau) = \cC_{P} (N; \tau) + \sigma \,\cC_{N\!P}(N; \tau)\,.\label{eq:TSgen}
\end{equation}
In this expression $\cC_{P} (N;\tau) $ contains the formal asymptotic perturbative expansion in $1/N$ for the integrated correlator $\corr$ given by
\begin{equation}\label{eq:originN}
 \cC_{P} (N;\tau) = \frac{N^2}{4}+ \sum_{\ell =0}^\infty N^{\frac{1}{2}-\ell} \sum_{m=0}^{\lfloor \ell/2\rfloor} \tilde{b}_{\ell ,m} \eisen{\tfrac{3}{2}+\delta_\ell +2m}\,,
\end{equation}
where $\delta_\ell=\ell \,(\rm{mod}\,2)$ and we denote with $\lfloor x\rfloor$ the floor of $x$. 
As discussed in \cite{Dorigoni:2022cua}, the constant coefficients $ \tilde{b}_{\ell ,m}$ can be easily computed starting from the generating series \eqref{eq:cGSU} but otherwise are not known in closed form for arbitrary $\ell$ and $m$.  The first few coefficients $ \tilde{b}_{\ell ,m}$ for $\ell\leq 4$ where already computed in  \cite{Chester:2019jas}, while expressions for general $\ell$ and fixed $m$ can be found in \cite{Dorigoni:2021guq}.

We stress that although this power-series in $1/N$ does not converge for any value of $\tau$, it is nonetheless manifestly a modular invariant function of $\tau$ order by order in $1/N$. The only $\tau$ dependence in \eqref{eq:originN} appears through the well-known modular invariant functions called \textit{non-holomorphic Eisenstein series}, which in the present convention are defined as
\begin{align}
  &\label{Eisen_integral_rep}  E^*(s;\tau) \coloneqq \frac{\Gamma(s)}{2}\sum_{(m,n)\neq (0,0)} \Ymn^{-s} = \frac{1}{2}\sum_{(m,n)\neq (0,0)}\int_0^\infty e^{-t\, \Ymn}\,t^{s-1}\,\dd t\,\\
&\notag =\xi(2s)\tau_2^s  +  \xi(2s-1)\, \tau_2^{1-s}+ \sum_{   k\neq0}e^{2\pi i k\tau_1}    2\sqrt{\tau_2}\,  |k|^{s-\half}\sigma_{1-2s} (k)     \, K_{s-\half}(2\pi |k| \tau_2) \,,
\end{align}
where $\xi(s)\coloneqq  \pi^{-s/2}\Gamma(s/2) \zeta(s) = \xi(1-s)$ denotes the completed Riemann zeta function, while $K_s(y)$ is a modified Bessel function and $\sigma_s(k) \coloneqq \sum_{d|k} d^s$ is the standard sum over positive divisors.
We note that while in the first line of \eqref{Eisen_integral_rep} both the lattice sum and the integral are only well defined for $\Re(s)>1$, the non-holomorphic Eisenstein series $E^*(s;\tau)$ can nonetheless be analytically continued to a meromorphic function of $s\in \mathbb{C}$ satisfying the functional equation $E^*(s;\tau) = E^*(1-s;\tau)$.

For the perturbative sector \eqref{eq:originN}, the coefficient of each order in $1/N$ is given by a finite sum of non-holomorphic Eisenstein series, $\eisen{s}$, of half-integer index $s$ ranging from a maximal value $s =\tfrac{3}{2}+\ell$ to a minimal value $s=\tfrac{3}{2}$. From  $\cC_{P} (N;\tau)$ we can recover  higher derivative corrections to the flat space-limit in the type IIB S-matrix of four gravitons at finite string coupling $\tau$ via the holographic dictionary \cite{Chester:2019jas}.

Importantly, the transseries expansion \eqref{eq:TSgen} of the full integrated correlator $\cC_{N}(\tau) $ does also contain non-perturbative, exponentially suppressed terms at large-$N$, captured by $\cC_{N\!P}(N;\tau)$ and given by the formal series
\begin{align} \label{eq:npSU}
\cC_{N\!P} (N;\tau)  &=    \sum_{\ell=0}^{\infty} N^{2-\frac{\ell}{2}} \sum_{m=0}^{\ell} \tilde{d}_{\ell,m} D_N\big(\tfrac{\ell}{2} -2m; \tau\big) \, ,
\end{align}
where the novel modular invariant function $D_N(s; \tau)$ is defined as\footnote{Note that compared to \cite{Dorigoni:2022cua}, where this class of functions was first introduced, we here use a slightly different and more convenient normalisation $D_N^{\rm{there}}(s;\tau) = 2^{4s}D_N^{\rm{here}}(s;\tau)$, which for the expansion  \eqref{eq:npSU}, where a different indexing is implemented, in turns implies $d^{\rm{there}}_{\ell,m}  = 2^{6\ell-8m}\tilde{d}_{\ell,\ell-m}$.}
\begin{equation}\label{Dfn_def}
    \Dfn{s}\coloneqq \sum_{(m,n)\neq (0,0)}\exp{\Big(-4\sqrt{N\Ymn}\Big)}\big(16 \Ymn\big)^{-s}\,.
\end{equation}
Via the holographic dictionary, it was conjectured in \cite{Dorigoni:2022cua} that these non-perturbative corrections capture the contribution of $\ell$ coincident $(p, q)$-string Euclidean world-sheet instantons wrapping a great two-sphere, $S^2$, on the equator of the five-sphere, $S^5$.

Finally, it was also argued that there is an ambiguity in resumming the large-$N$ asymptotic perturbative expansion \eqref{eq:originN} which has to be compensated by a change in the non-perturbative sector captured by $\cC_{N\!P}(N;\tau)$. This amounts to a jump in the transseries parameter $\sigma$, which is a piece-wise constant function of $\arg(N)$, taking values $\sigma = \pm\,i$ according to $\arg(N) >0$ or $<0$ respectively.
In section \ref{resurgence_section}, we show that the non-perturbative corrections \eqref{eq:npSU}, as well as the transseries parameter $\sigma$, can be fixed completely from a proper resurgence analysis of a modified modular invariant Borel resummation of the purely perturbative data \eqref{eq:originN}. This explains the resurgent origins of the transseries \eqref{eq:TSgen}, originally found solely via generating series methods. 

\subsection{Spectral representation}
\label{sec:spec}

An alternative and equivalent representation to the lattice-sum integral expression \eqref{corr_int_definition} is obtained via $\rm{SL}(2,\mathbb{Z})$ spectral theory, a method of decomposing any (suitable) modular invariant function as a linear combination of “good” basis elements, i.e. $L^2$-normalisable eigenfunctions of the hyperbolic Laplace operator $\Delta_\tau$.

For the present discussion we only highlight the fact that this integrated correlator possesses an astonishingly simple spectral representation which only involves an integral over special normalisable eigenfunctions: the non-holomorphic Eisenstein series $E^*(s;\tau)$ with $\mbox{Re}(s) = \tfrac{1}{2}$.  
The coefficient of a given non-holomorphic Eisenstein series in this linear combination is called the \textit{spectral overlap}. This spectral approach to the integrated correlator \eqref{eq:GHdef} was first derived in \cite{Collier:2022emf} where the interested reader can also find more details on spectral theory. 
 
Here we rederive the spectral representation for $\corr$ starting directly from the lattice-sum integral representation \eqref{corr_int_definition}. 
We begin by considering the lattice sum in \eqref{corr_int_definition} and splitting it into the sum of the $(m,n)=(0,0)$ contribution and terms with $(m,n)\neq (0,0)$, 
\begin{equation}\label{correlator_split}
    \corr = \frac{1}{2}\int_0^\infty B_N(t)\,\dd t + \frac{1}{2}\sum_{(m,n)\neq (0,0)}\int_0^\infty e^{-t\,\Ymn}\,B_N(t)\,\dd t\,.
\end{equation}

We now rewrite the second term in this expression in terms of the Mellin transform, $M_N(s)$, of the function $B_N(t)$ defined as
\begin{equation} 
M_N(s)\coloneqq\int_0^\infty t^{s-1}\,B_N(t) \,\dd t\,.\label{eq:Mellin}
\end{equation}
Given the expression \eqref{eq:BSUN} for $B_N(t)$, this Mellin integral can be shown to converge in the strip $-1<\mbox{Re}(s)<2$ and has an analytic continuation to a meromorphic function of $s\in \mathbb{C}$.
This transform can be inverted via Mellin inversion formula,
\begin{equation}\label{mellin_inversion}
    B_N(t) = \int_{\Re(s)=\alpha} t^{-s} M_N(s)\,\frac{\dd s}{2\pi i},
\end{equation}
where the constant $\alpha\in\mathbb{R}$ is chosen in such a way that the original Mellin integral \eqref{eq:Mellin} converges for $\Re(s)=\alpha$.
Crucially, we notice that the functional equation \eqref{B_inversion_identity} translates immediately to the reflection formula,
\begin{equation}
M_N(1-s)=M_N(s)\,. \label{eq:RefMell}
\end{equation} 

We now substitute Mellin inversion formula in \eqref{correlator_split} and perform a reflection $s\to 1-s$ while using \eqref{eq:RefMell} to arrive at,
\begin{equation}
    \corr = \frac{1}{2}\int_0^\infty B_N(t)\,\dd t + \int_{\Re(s)=1+\epsilon} M_N(s)\left( \frac{1}{2}\sum_{(m,n)\neq (0,0)}\int_0^\infty e^{-t\,\Ymn}t^{s-1}\dd t  \right) \frac{\dd s}{2\pi i}\,.
\end{equation}
The $s$-contour of integration has been shifted to $\Re(s)=1+\epsilon$, with $\epsilon>0$ sufficiently small, so that the $t$-integral and the lattice-sum are both convergent and we are allowed to use the integral representation \eqref{Eisen_integral_rep} for the non-holomorphic Eisenstein series, thus arriving at the sought-after spectral representation for the integrated correlator 
\begin{align}\label{C_spec_rep}
    \corr &\,= \langle\mathcal{C}_N\rangle +\int_{\Re(s)=\frac{1}{2}} M_N(s)E^*(s;\tau)\frac{\dd s}{2\pi i}\,,\\
     \langle\mathcal{C}_N\rangle  &\coloneqq  \int_0^\infty B_N(t)\,\dd t  =\lim_{s\to 1} M_N(s)\,,\label{eq:avg}
\end{align}
where the additional factor $1/2$ for the constant term originates from having moved the contour of integration back to $\Re(s)=\frac{1}{2}$ combined with the fact that $\textrm{res}_{s=1}E^*(s;\tau) = \frac{1}{2}$.

While in the lattice-sum representation the $N$ dependence is encoded entirely in the rational functions $B_N(t)$ given in \eqref{eq:BSUN}, here this information is captured by the spectral overlap function, i.e. the Mellin transform $M_N(s)$.
The function $M_N(s)$ can be obtained by exploiting an intriguing Laplace-difference equation found in \cite{Dorigoni:2021guq} and satisfied by the integrated correlator:
\begin{align}\label{recursion}
    \notag\Delta_\tau \correl{N} - &(N^2-1)\Big(\correl{N+1}-2\correl{N}+\correl{N-1}\Big) \\&- (N+1)\correl{N-1}+(N-1)\correl{N+1}=0,
\end{align}
which fixes $\correl{N}$ in terms of the initial data $\correl{2}$ and $\correl{1}=0$. 

By specialising \eqref{eq:BSUN} to the $SU(2)$ theory, i.e. by setting $N=2$, we obtain the initial condition, 
\begin{equation}
    B_2(t) = \frac{9t-30t^2+9t^3}{(t+1)^5}\,,
\end{equation}
from which it is immediate to derive its Mellin transform,
\begin{equation}
   \qquad M_2(s) = \frac{\pi s(1-s)(2s-1)^2}{2 \sin{(\pi s)}}\,.
\end{equation}
We can then combine \eqref{C_spec_rep} with \eqref{recursion} and the known Laplace equation 
\begin{equation}
\Delta_\tau E^*(s;\tau) =s(s-1) E^*(s;\tau)\,,
\end{equation}
to find a recurrence relation satisfied by the spectral overlaps, namely
\begin{equation}
    N(N-1)M_{N+1}(s) = [s(s-1)+2(N^2-1)]M_N(s)- N(N+1)M_{N-1}(s)\,.\label{eq:LapDiffSpec}
\end{equation}
As show in \cite{Paul:2023rka}, this recursion is solved by
\begin{align}\label{MellinN_formula}
    M_N(s)=&\frac{N(N-1)}{4}\frac{\pi s(1-s)(2s-1)^2}{\sin{(\pi s)}} {}_3F_2(2-N,s,1-s;3,2\vert 1)\,,
\end{align}
thus implying from \eqref{eq:avg} that $\average{\mathcal{C}_N}={N(N-1)}/{4}$.
The hypergeometric function in this equation is somewhat misleading, since the parameter $2-N$ is a non-positive integer for $N\geq2$ and as a consequence, the hypergeometric function always reduces to a polynomial in $s(1-s)$ of degree $N-2$.
In appendix \ref{spectral_rep_app} we find a more convenient expression given by
\begin{equation}\label{new_expr_spect_rep}
     M_N(s) =   \frac{2^{-2s} (2 s-1) \Gamma \left(\frac{3}{2}-s\right)}{\sqrt{\pi } \,\Gamma (-s)}\int_0^1 x^{s-3}(1-x)^N \tFo{s-1}{s}{2s}{x}\dd x + (s\leftrightarrow 1-s)\,,
\end{equation}
where a certain regularisation is required when treating the $x$-integral near $x= 0$, see in particular equation \eqref{I_regularisation} and the detailed analysis presented in appendix \ref{spectral_rep_app}.

While at finite $N$ it is straightforward to evaluate the Mellin transform  \eqref{MellinN_formula} and obtain the spectral representation for the integrated correlator \eqref{C_spec_rep}, it is absolutely not obvious how to deduce its large-$N$ expansion. In particular, prior to the present work, it has not been shown how to reconstruct the complete transseries expansion \eqref{eq:TSgen} starting directly from the spectral decomposition \eqref{C_spec_rep}. 
In section \ref{spectral_section} we start from the spectral overlap \eqref{new_expr_spect_rep} to manifest how the resurgent structure of the integrated correlator is beautifully encoded in the spectral representation \eqref{C_spec_rep}.

\section{Resurgence of modular invariant transseries}
\label{resurgence_section}

%

In this section we want to show how the exact large-$N$ transseries expansion at fixed $\tau$ for $ \cC_{N}(\tau)$ displayed in \eqref{eq:TSgen}, so far only analysed numerically in \cite{Hatsuda:2022enx} and derived \cite{Dorigoni:2022cua} via generating series methods, can be derived using resurgence analysis. In particular, we prove that it is possible to reconstruct the non-perturbative and modular invariant contributions \eqref{eq:npSU} from a suitable resummation of the large-$N$ formal, yet modular invariant perturbative sector~\eqref{eq:originN}.

We start by focusing our attention on the purely perturbative expansion of the integrated correlator \eqref{eq:GHdef}, which at large-$N$ and fixed $\tau$ has the formal asymptotic perturbative expansion \eqref{eq:originN}, here rewritten for convenience
\begin{equation}\label{eq:originN1}
 \cC_{N}(\tau)  \sim \cC_{P} (N;\tau) = \frac{N^2}{4}+ \sum_{\ell =0}^\infty N^{\frac{1}{2}-\ell} \sum_{m=0}^{\lfloor \ell/2\rfloor} \tilde{b}_{\ell ,m} \eisen{\tfrac{3}{2}+\delta_\ell +2m}\,,
\end{equation}
with $\delta_\ell=0$ for even $\ell$ and $\delta_\ell=1$ for odd $\ell$.

As already noted previously, in the perturbative sector the coefficient of each order in $1/N$ is given by a finite sum of non-holomorphic Eisenstein series, $\eisen{s}$, of half-integer index $s$ ranging from the maximal value $s =\tfrac{3}{2}+\ell$ to the minimal one $s=\tfrac{3}{2}$. Following \cite{Alday:2023pet}, we reorganise this formal power series in $1/N$ as  
\begin{equation}\label{C_pert_exp}
     \cC_{P} (N;\tau)  = \frac{N^2}{4}+\sum_{r=0}^\infty N^{2-2r} \PhiP{r}\,,
\end{equation}
having defined
\begin{equation}\label{C_pert_expPhiR}
      \PhiP{r} \coloneqq \sum_{k=0}^\infty b_{r,k} N^{-\frac{3}{2}-k} \eisen{\tfrac{3}{2}+k}\,.
\end{equation}
Here we made the change of summation variables $\ell =2r+k$ and $m = \lfloor \tfrac{k}{2} \rfloor$ and correspondingly denoted the rearranged coefficients by $b_{r,k} = \tilde{b}_{\ell,m}$.
For fixed $r$, the formal power-series $ \PhiP{r}$ can be understood as collecting, order by order in $1/N$, the contributions to \eqref{eq:originN1}  coming from the ``$r$-subleading index'' non-holomorphic Eisenstein series, i.e. all $\eisen{s}$ with index $s=\tfrac{3}{2}+\ell -2 r$.

For example, we can focus on the contribution to \eqref{eq:originN1} coming only from grouping all ``leading-index'' non-holomorphic Eisenstein series, i.e. all terms in \eqref{eq:originN1} with $m=\lfloor \ell/2\rfloor$ or equivalently specialising \eqref{C_pert_expPhiR} to $ k=\ell$ and $r=0$:
\begin{align}
\PhiP{0} =\sum_{k=0}^\infty b_{0,k}N^{-\frac{3}{2}-k} \eisen{\tfrac{3}{2}+k} \,,\label{eq:PhiP0}
\end{align}
where the coefficients $b_{0,k} = \tilde{b}_{k ,\lfloor k/2\rfloor}  $  have been computed in \cite{Dorigoni:2021guq} and are given by
\begin{equation}
 b_{0,k} \coloneqq  \frac{(k+1)\Gamma(k-\frac{1}{2})  \Gamma(k+\frac{5}{2})}{2^{2k+1} \,\pi^{3/2}\, \Gamma(k+1)} \,.\label{eq:b0k}
\end{equation}

As manifest from this particular example (and other cases presented in \cite{Dorigoni:2021guq,Dorigoni:2022cua}), we notice that the coefficients $b_{r,k}$ appearing in the series \eqref{C_pert_expPhiR} grow factorially with $k$ for fixed $r$, i.e. $b_{r,k}\sim k!$, so that $\PhiP{r}$ can be thought of as a formal asymptotic series with coefficients given by rational multiples of half-integer non-holomorphic Eisenstein series. 
This simple observation suggests immediately that a proper Borel-like resummation of the formal perturbative expansion \eqref{C_pert_expPhiR}, and hence of the whole perturbative sector \eqref{C_pert_exp}, should by consistency require the introduction of the anticipated non-perturbative terms $\cC_{N\!P} (N;\tau)$ presented in \eqref{eq:npSU} and here rewritten for convenience,
\begin{align} \label{eq:npSU1}
\cC_{N\!P} (N;\tau)  &=    \sum_{\ell=0}^{\infty} N^{2-\frac{\ell}{2}} \sum_{m=0}^{\ell} \tilde{d}_{\ell,m} D_N\big(\tfrac{\ell}{2} -2m; \tau\big) \, .
\end{align}
We stress once more that in \cite{Dorigoni:2022cua} these terms have been recovered starting from the generating series \eqref{eq:gresult}, while presently we are in the process of describing how to retrieve them from a resurgence analysis approach to the resummation of the perturbative sector \eqref{C_pert_expPhiR}.

To this end, we proceed just like we did in the perturbative sector starting from \eqref{eq:originN1} to arrive at \eqref{C_pert_exp}, and rearrange the non-perturbative terms \eqref{eq:npSU1} as  
\begin{align}
\cC_{N\!P} (N;\tau)  &=  \sum_{r=0}^\infty N^{2-2r}  \PhiNP{r} \,,\\
 \PhiNP{r}  &\label{eq:NPdrk}= \sum_{k=-3r-1}^\infty d_{r,k}\,N^{-\frac{k+1}{2}}   \Dfn{\tfrac{k+1}{2}} \,,
\end{align}
where we made the change of summation variables $\ell =4r+k+1$ and $m =r$ and correspondingly denoted $d_{r,k} = \tilde{d}_{\ell,m}$.
Although using the methods of \cite{Dorigoni:2022cua}\footnote{\label{foot}We note again that, due to the change in normalisation \eqref{Dfn_def} and  in summation variables, to compare the non-perturbative coefficients $d_{r,k}$ with the results of \cite{Dorigoni:2022cua} we must use $d^{\rm{there}}_{r,k}  ={2^{6 r-8 k}} d_{r-k,4 k-3 r-1}$.} it is possible to compute the coefficients $d_{r,k}$ for different values of $r$ and $k$, no analytic expression similar to \eqref{eq:b0k} has been found prior to this work. Using resurgence analysis we show how to derive the coefficients $d_{r,k}$ from the perturbative coefficients $b_{r,k}$ and manifest that at fixed value of $r$ the numbers $d_{r,k}$ are once again factorially divergent as $k\to\infty$.

In what follows we show that the large-$N$ transseries representation~\eqref{eq:TSgen} for the integrated correlator~\eqref{eq:GHdef} can be recovered from the Borel-Écalle median resummation of the perturbative sectors~\eqref{C_pert_expPhiR},~i.e.
\begin{align}
 \cC_{N}(\tau) &\label{C_full_transseries} =\frac{N^2}{4}+ \sum_{r=0}^\infty N^{2-2r}  \mathcal{C}^{(r)}(N;\tau) \,,\\
    \mathcal{C}^{(r)}(N;\tau) &\label{eq:CrTS}= \PhiP{r}+\sigma\,\PhiNP{r}\,.
\end{align}
As already mentioned previously, the median resummation contains an additional parameter called the \textit{transseries parameter} $\sigma = \sigma(\mbox{arg}(N) )$ which is a piecewise constant function of $\mbox{arg}(N)$.
We will show that for the median resummation here considered the transseries parameter takes values $\sigma= \pm\, i$ according to whether $\mbox{arg}(N)>0$ or $<0$, this will in turn be correlated with the how we perform the resummation of the perturbative sector $ \PhiP{r}$.

Thanks to our modular invariant resurgence analysis approach we find that:
\begin{enumerate}
\item[(i)] From the ``$r$-subleading index'' non-holomorphic Eisenstein series  $\eisen{s}$ with index $s=\tfrac{3}{2}+\ell - {2r}$, grouped in $\PhiP{r} $, we can retrieve all of the ``$r$-subleading index'' non-perturbative terms $ \Dfn{s}$ with $s= \tfrac{\ell}{2}-2r$, grouped in $ \PhiNP{r} $, see section \ref{sec:ResIC};
\item[(ii)] As a consequence of modularity, in the 't Hooft limit where $\lambda = \sqrt{4 \pi N/\tau_2} $ is kept fixed, the function $\mathcal{C}^{(r)}(N;\tau)$ reduces to the transseries expansion of the genus-$r$ contribution to the integrated correlator, as well as the transseries expansion of the ``dual 't Hooft-limit'' genus-$r$ contribution where $\tilde{\lambda} \coloneqq (4\pi N)^2/ \lambda$ is kept fixed, see section \ref{tHooft_section};
\item[(iii)] The sum over $r$ in \eqref{C_full_transseries} is actually Borel summable and does not introduce any additional non-perturbative corrections. Furthermore, the large-$N$ expansion of the spectral representation \eqref{C_spec_rep}  leads directly to the spectral representation of $\mathcal{C}^{(r)}(N;\tau)$ whose spectral overlap encodes quite naturally both the perturbative, $\PhiP{r}$, and non-perturbative, $\PhiNP{r}$, sectors; see section \ref{spectral_section}
\end{enumerate}

\subsection{Modular invariant resummation at large-$N$}
\label{sec:resum}

Motivated from the case of present interest, namely the formal perturbative series \eqref{C_pert_expPhiR}, the main goal of this section is to define a modular invariant resummation for the formal but modular invariant series,
\begin{equation}\label{original_series}
    \Phi_P(N;\tau)\coloneqq \sum_{k=0}^\infty b_k\, N^{-\frac{3}{2}-k}   \eisen{\tfrac{3}{2}+k}\,,
\end{equation}
where the coefficients $b_k$ diverge factorially fast, i.e. $b_k\sim k!$.

 Inspired by the particular exponential structure of the candidate non-perturbative terms, $\Dfn{s}$, defined in \eqref{Dfn_def}, we introduce a somewhat non-standard integral representation for the non-holomorphic Eisenstein series 
 \begin{equation}\label{Eisen_integr_rep}
    N^{-s}\eisen{s} = \int_0^\infty \NPfn{\sqrt{N}t}\, \frac{2 \Gamma(s)}{\Gamma(2s)}(4t)^{2s-1} \dd t\,,
\end{equation}
 where we have defined a modular invariant modified Borel kernel
 \begin{equation}\label{NPfn_def}
    \NPfn{t} \coloneqq  D_{t^2}(0;\tau)  = \summn e^{-4t\sqrt{\Ymn}} \,,
\end{equation}
which converges absolutely for all $\tau$ in the upper-half plane  when $\mbox{Re}(t)>0$. Some of the properties of $ \NPfn{t}$ are presented in appendix \ref{sec:AppE}, in particular from \eqref{NPfn_spect_rep} we see that  \eqref{Eisen_integr_rep} is convergent for $\mbox{Re}(s)>{1}$.
%

 We are now in a position to define the Borel transform of the formal series \eqref{original_series} as
\begin{equation}
    \mathcal{B}[\Phi_P](t) \coloneqq \sum_{k=0}^\infty b_k\,\frac{2\Gamma(k+\frac{3}{2})}{\Gamma(2k+3)}(4t)^{2k+2}\,,\label{eq:BorelDef}
\end{equation}
which has a positive radius of convergence in the complex Borel $t$-plane under the assumption that $b_k\sim k!$, thus defining a germ of analytic functions at the origin. 

Following standard resurgence analysis arguments, see e.g. the introductory notes \cite{Dorigoni:2014hea}, we combine \eqref{eq:BorelDef} with the integral representation \eqref{Eisen_integr_rep} specialised to $s=k+\frac{3}{2}$, and define the \textit{directional Borel resummation} of the original formal series \eqref{original_series} as
\begin{equation}\label{resummation_def}
    \mathcal{S}_\theta [\Phi_P](N;\tau) \coloneqq \int_0^{e^{i\theta}\infty} \NPfn{\sqrt{N}t}  \mathcal{B}[\Phi_P](t)\dd t\,.
\end{equation}
If the direction of integration $-\pi< \theta \leq \pi$ is such that the Borel transform $\mathcal{B}[\Phi_P](t)$ has no singularities, i.e. if $\mbox{arg} (t) = \theta$ is \textit{not} a Stokes direction, we have that the directional Borel resummation \eqref{resummation_def} is well-defined (under a moderate growth condition for the Borel transform) and it defines a modular invariant function of $\tau$, which is analytic in $N$ in the wedge $\mbox{Re}(\sqrt{N} e^{i\theta})>0$ of the complex $N$-plane. From equation \eqref{Eisen_integr_rep}, we see that the asymptotic expansion of \eqref{resummation_def} at large-$N$ reproduces the formal expansion \eqref{original_series} we started with, i.e. we have resummed \eqref{original_series} in a modular invariant way.

Furthermore, given two directions $\theta_1$ and $\theta_2$ with $\theta_1<\theta_2$ such that $\mathcal{B}[\Phi_P](t)$ is regular in the wedge $\theta_1\leq \mbox{arg}(t) \leq \theta_2$, we find that $ \mathcal{S}_{\theta_1} [\Phi_P](N;\tau) = \mathcal{S}_{\theta_2 }[\Phi_P](N;\tau)$ on the common domain of analyticity. Hence $ \mathcal{S}_{\theta_2 }[\Phi_P](N;\tau)$ defines an analytic continuation of $ \mathcal{S}_{\theta_1 }[\Phi_P](N;\tau)$ to a wider wedge of the complex $N$-plane. 
However, if the direction $\theta= \theta_\star$ is a singular direction for  $\mathcal{B}[\Phi_P](t)$, usually called a \textit{Stokes direction}, we find instead that the analytic functions $ \mathcal{S}_{\theta_\star-\epsilon} [\Phi_P](N;\tau)$ and $ \mathcal{S}_{\theta_\star+\epsilon} [\Phi_P](N;\tau)$, with $\epsilon\to0^+$, do not coincide on the common domain of analyticity and they crucially differ by non-perturbative terms.
Near a Stokes direction $\theta_\star$, we are then naturally led to consider the \textit{lateral Borel resummations} defined as
\begin{equation}
\mathcal{S}_{\theta_\star^\pm} [\Phi_P](N;\tau) \coloneqq \lim_{\epsilon\to0^+} \mathcal{S}_{\theta_\star\pm \epsilon} [\Phi_P](N;\tau)\,.
\end{equation}

In the case where $N$ denotes the number of colours, we obviously want to define a resummation of \eqref{original_series} which is analytic in a neighbourhood of the positive real axis $\mbox{arg}(N)=0$. However, we will shortly see that for the cases of interest the direction $\theta=0$ happens to be a Stokes ray. In particular, we need to consider the case where the Borel transform $\mathcal{B}[\Phi_P](t)$ has polar singularities at $t=1$ plus a branch cut starting at $t=1$ with an expansion of the form 
\begin{equation}\label{sing_t1}
    \mathcal{B}[\Phi_P](t) \sim- \frac{1}{\pi}\sum_{k=1}^{M}  \frac{d_{-k} (k-1)!}{ (1-t)^{k}} + \Big(\sum_{k=0}^\infty \frac{d_k (t-1)^k}{k!} \Big)\frac{\log{(1-t)}}{\pi} +\text{reg}(t-1),
\end{equation}
with $M$ a positive integer specifying the maximal order of the pole, while $\text{reg}(t-1)$ denotes the analytic part at $t=1$. Besides the polar part, we also have a logarithmic singularity multiplied by a new germ of analytic functions at the origin, which is specified by a series of factorially divergent coefficients $d_k$.

Starting from \eqref{sing_t1}, we easily compute the difference between the two lateral resummations of the original series \eqref{original_series}, related to the so-called Stokes automorphism, which takes the form
\begin{equation}\label{discont}
    \Big(\mathcal{S}_+-\mathcal{S}_-\Big)[\Phi_P](N;\tau) = -2i\sum_{k=-M}^\infty d_k\, N^{-\frac{k+1}{2}}  \Dfn{\tfrac{1+k}{2}}\,,
\end{equation}
where for ease of notation we write $\mathcal{S}_\pm  \coloneqq \mathcal{S}_{0^\pm}$ since $\theta = 0$ will be the only Stokes line we need considering.
Notice that since the coefficients $d_k$ are in general factorially divergent, as we show for the integrated correlator, the discontinuity equation \eqref{discont} defines once again a formal series, which is however modular invariant and whose coefficients are no longer given by non-holormorphic Eisenstein series but rather they belong to the class of functions defined in \eqref{Dfn_def}. 

We stress how general this result is: given a formal series of the form \eqref{original_series} whose Borel transform, $\mathcal{B}[\Phi_P](t)$, defines a germ of analytic function at the origin with a singular structure akin to \eqref{sing_t1}, along any Stokes direction we must have a non-perturbative and modular invariant discontinuity in lateral resummations  captured by an infinite sum of $D_N(s;\tau)$ functions. 

To finally construct a complete modular invariant transseries starting from the purely perturbative sector \eqref{original_series}, we additionally need to understand how to modify the lateral resummations $\mathcal{S}_\pm$ to take into account their discontinuity \eqref{discont}. Proceeding as we just did for the perturbative sector, we start from the non-perturbative modular invariant series in \eqref{discont}
\begin{equation}\label{PhiNP_def}
    \Phi_{N\!P}(N;\tau) \coloneqq \sum_{k=-M}^\infty  d_k\,N^{-\frac{k+1}{2}}  \Dfn{\tfrac{1+k}{2}}\,,
\end{equation} 
and consider the integral representation 
\begin{align}
    N^{-\frac{k+1}{2}}\Dfn{\tfrac{k+1}{2}} & =\label{DN_int_rep}\int_0^\infty \NPfn{\sqrt{N}(t+1)}  \frac{t^{k}}{\Gamma(k+1)}\dd t\,,
\end{align}
with~$k\geq 0 $.
Just like the integral representation \eqref{Eisen_integr_rep} leads to the perturbative Borel transform \eqref{eq:BorelDef}, we now use \eqref{DN_int_rep} to define the directional Borel resummation of the non-perturbative sector~\eqref{PhiNP_def},
\begin{align}
    &\widetilde{\mathcal{B}}[\Phi_{N\!P}](t) \coloneqq \sum_{k=0}^\infty \frac{d_k}{\Gamma(k+1)}t^{k},\\ \label{resummation_NP_def}
    &\mathcal{S}_\theta[\Phi_{N\!P}](N;\tau) \coloneqq \sum_{k=-M}^{-1}  d_k\,N^{-\frac{k+1}{2}} \Dfn{\tfrac{1+k}{2}}+\int_0^{e^{i\theta}\infty}\NPfn{\sqrt{N}(t+1)} \,\widetilde{\mathcal{B}}[\Phi_{N\!P}](t) \,\dd t\,.
\end{align}
Note that the finitely many terms in \eqref{resummation_NP_def} with a positive power of $N$ have to be treated separately since they cannot be represented via \eqref{DN_int_rep}, these terms correspond to the polar part of the singular behaviour of the Borel transform \eqref{sing_t1}. 

In standard applications of resurgence theory the Borel kernel is given by the usual Laplace measure $e^{- \sqrt{N}t }\dd t$. In this case it is well appreciated that under a shift of integration variable $t\to t+1$, the measure naturally factors into the same integration kernel multiplied by the expected exponential suppression factor $^{-\sqrt{N}}$ which characterises the non-perturbative sectors. 
Due to the lattice sum nature \eqref{NPfn_def} of the present modular invariant Borel integration kernel $\NPfn{\sqrt{N}t}$, we do not have such property, i.e. 
$$\NPfn{\sqrt{N}(t+1)}\dd t \neq \NPfn{\sqrt{N}} \, \NPfn{\sqrt{N}t} \dd t\,.$$
This explains why we have to define a second Borel transform $\widetilde{\mathcal{B}}[\Phi_{N\!P}](t)$ in \eqref{resummation_NP_def}: while the building blocks of the perturbative sector are given by non-holomorphic Eisenstein series, they differ from those of the non-perturbative part,~i.e. the functions $\Dfn{s}$. However, our modular invariant Borel kernel contains both objects:
\begin{align}
N^{-s}\eisen{s} &\notag = \int_0^\infty \NPfn{\sqrt{N}t}\, \frac{2 \Gamma(s)}{\Gamma(2s)}(4t)^{2s-1} \dd t\,,\\
\qquad N^{-s}\Dfn{s} &= \int_0^\infty \NPfn{\sqrt{N}(t+1)}\,\frac{t^{2 s-1}}{\Gamma (2 s)} \dd t\, .
\end{align}

Given the discontinuity \eqref{discont}, it is now manifest that for the physically relevant domain $N>0$, the two resummations $\mathcal{S}_\pm [\Phi_P](N;\tau)$ do differ and we have an ambiguity in how we resum the purely perturbative formal power series \eqref{original_series}.
Furthermore, while the original formal power series \eqref{original_series} is manifestly real for $N>0$ and $\tau \in \mathfrak{H}$, neither of the two lateral resummations $\mathcal{S}_\pm [\Phi_P](N;\tau)$ is. 
To obtain a real and unambiguous resummation for $N>0$, we have to consider an average
between the two lateral resummations $\mathcal{S}_\pm [\Phi_P](N;\tau)$, usually referred to as median resummation  \cite{delabaere1999resurgent},
\begin{equation}
    \Phi(N;\tau) = \Phi_P(N;\tau) + \sigma\, \Phi_{N\!P}(N;\tau)\,.
\end{equation}
The additional parameter $\sigma$, called the transseries parameter, is the piece-wise constant function of $\mbox{arg}(N)$ given by $\sigma = \pm i $ according to $\mbox{arg}(N){\gtrless}0$, which in turn is correlated with the choice of lateral resummation 
\begin{equation}\label{median_resum_def}
    \mathcal{S}_{med}[\Phi_P](N;\tau) \coloneqq \left\lbrace \begin{matrix}
     \mathcal{S}_{+}[\Phi_P] (N;\tau)+ i\, \mathcal{S}_0[ \Phi_{N\!P}](N;\tau) \,,\qquad\qquad \mbox{arg}(N)>0\,,\\
      \mathcal{S}_{-}[\Phi_P](N;\tau)-i\,\mathcal{S}_0[\Phi_{N\!P}](N;\tau)\,,\qquad\qquad \mbox{arg}(N)<0\,.
      \end{matrix}\right.
\end{equation}
We shortly show that the equality between the two seemingly different expressions comes from the discontinuity equation \eqref{discont} combined with the fact that for the integrated correlator $\mbox{arg}(N)=0$ is not a Stokes direction for $\Phi_{N\!P}(N;\tau)$, which can then be resummed via \eqref{resummation_NP_def} along $\theta=0$, i.e. by considering $ \mathcal{S}_0[ \Phi_{N\!P}](N;\tau)$.

The median resummation produces an unambiguous resummation of~\eqref{original_series} in a wedge of the complex-$N$ plane which contains the physical domain $N>0$. In particular, it is easy to show that~\eqref{median_resum_def} is real-analytic for~$N>0$ and~$\tau \in \mathfrak{H}$, since it can be rewritten as
\begin{align}
    \mathcal{S}_{med}[\Phi_P](N;\tau)&\label{eq:Smed} = \frac{1}{2}(\mathcal{S}_++\mathcal{S}_-)[\Phi_P](N;\tau) = \intmed\NPfn{\sqrt{N}t}  \mathcal{B}[\Phi_P](t)\dd t\,,\\
    &\notag = \int_0^\infty \NPfn{\sqrt{N}t}  \mbox{Re}\Big(\mathcal{B}[\Phi_P](t) \Big)\dd t\,.
\end{align}
Here we have defined for later convenience the notation for the median integration
\begin{equation}\label{intmed_def}
    \intmed \coloneqq \frac{1}{2}\Big(\int_0^{\infty+i\epsilon} + \int_0^{\infty-i\epsilon} \Big)\,,
\end{equation} 
in the limit $\epsilon \to 0^+$.~The particular integral representation \eqref{eq:Smed} for the non-perturbative integrated correlator will be obtained directly from spectral theory in section \ref{spectral_section}.

\subsection{Resurgence of the integrated correlators}
\label{sec:ResIC}

Thanks to the analysis of the previous section, we have thus constructed an unambiguous resummation for a formal perturbative series in non-holomorphic Eisenstein series $\eisen{s}$, schematically presented in \eqref{original_series}, and have shown that it generically requires exponentially suppressed terms which involve the modular invariant function $\Dfn{s}$. 

In this section we show how an application of this resummation method to the two integrated correlators \eqref{eq:GHdef} yields the complete large-$N$ expansion of the first integrated correlator $\corr$, as well as a particular non-perturbative sector of the second integrated correlator $\mathcal{H}_N(\tau)$.
This explains the appearance of such non-perturbative terms in the full transseries representation \eqref{C_full_transseries} of the integrated correlator $\corr$ and our analysis establishes a connection between the perturbative coefficients $\tilde{b}_{\ell,m}$ in \eqref{eq:originN1}, or alternatively the coefficients ${b}_{r,k}$ in \eqref{C_pert_expPhiR}, and the non-perturbative coefficients $\tilde{d}_{\ell,m}$ in \eqref{eq:npSU1}, or alternatively the coefficients $d_{r,k}$ presented in \eqref{eq:NPdrk}, as we now show in more detail. 

\subsubsection*{First integrated correlator}

For concreteness we discuss the cases $r=0$ and $r=1$, although our analysis can be extended straightforwardly to arbitrary $r$. We begin by deriving the non-perturbative resummation of~$\mathcal{C}^{(0)}_P(N;\tau)$ presented in~\eqref{eq:PhiP0}, which contains all large-$N$ perturbative contributions originating from non-holomorphic Eisenstein series with leading index.

Given the definition \eqref{eq:BorelDef} we compute the associated Borel transform of this series, which takes the form
\begin{equation}\label{Phi0_Borel}
    \mathcal{B}[\mathcal{C}^{(0)}_P](t) = \frac{4}{\pi}\sum_{k=0}^\infty \frac{\Gamma(k-\frac{1}{2})\Gamma(k+\frac{5}{2})}{\Gamma(k+1)^2}t^{2k+2} = -6t^2 \tFo{-\tfrac{1}{2}}{\tfrac{5}{2}}{1}{t^2}.
\end{equation}
As anticipated, we see that the Borel transform has two Stokes directions: one for $\theta=0$ and the other for $\theta=\pi$, both with logarithmic branch cuts starting respectively at $t=\pm 1$ due to the hypergeometric function $_2F_1$.
Since we are interested in obtaining a non-perturbative resummation for $\mbox{arg}(N)=0$, we have to compute the singular behaviour of the Borel transform \eqref{Phi0_Borel} near the point $t=1$. This can be obtained from the integral representation of the hypergeometric function thanks to which we find 
\begin{equation}\label{Phi0_disc}
    \mathcal{B}[\mathcal{C}^{(0)}_P](t) \sim -\frac{(-2)}{\pi (1-t)} -9\,t^2 \tFo{-\tfrac{1}{2}}{\tfrac{5}{2}}{2}{1-t^2} \frac{\log{(1-t)}}{\pi} + \text{reg}(t-1)\,,
\end{equation}
where again $\text{reg}(t-1)$ denotes the analytic part at $t=1$.

This structure is precisely of the form \eqref{sing_t1} previously analysed and, as a consequence, we have that the two lateral resummations of \eqref{Phi0_Borel} do not coincide on the common domain of analyticity.
We can compute the difference in lateral resummations \eqref{discont} directly from the singular behaviour \eqref{Phi0_disc},
\begin{align}
 &   (\mathcal{S}_+-\mathcal{S}_-)[\mathcal{C}^{(0)}_P](N;\tau)  = -2 i \mathcal{S}_0[\mathcal{C}^{(0)}_{N\!P}](N;\tau)\\
 &\notag = 4i \Dfn{0}+18i\int_0^\infty  \NPfn{\sqrt{N}(t+1)}\,(1+t)^2\tFo{-\tfrac{1}{2}}{\tfrac{5}{2}}{2}{-t(2+t)} \dd t\,.
\end{align}

Following our general discussion, we use this discontinuity and \eqref{resummation_NP_def} to define the resummation of the
non-perturbative sector which is then given by
\begin{align}
&\mathcal{S}_0[\mathcal{C}^{(0)}_{N\!P}](N;\tau) \label{eq:NPC0}\\
&\notag = -2 \Dfn{0}-9\int_0^\infty  \NPfn{\sqrt{N}(t+1)}(1+t)^2\tFo{-\tfrac{1}{2}}{\tfrac{5}{2}}{2}{-t(2+t)} \dd t\,.
\end{align}
As already mentioned, the Borel transform of the non-perturbative sector is regular along the direction $\mbox{arg}(t)=0$, hence \eqref{eq:NPC0} is precisely the Borel resummation of the formal series of non-perturbative corrections
\begin{align}\label{Phi0_generating_series}
    \mathcal{C}^{(0)}_{N\!P}(N;\tau) &= \sum_{k=-1}^\infty d_{0,k}\,N^{-\frac{k+1}{2}} \Dfn{\tfrac{k+1}{2}} \\
    &\notag = -2\Dfn{0}-9N^{-\frac{1}{2}}\Dfn{\tfrac{1}{2}}-\frac{117}{4}N^{-1}\Dfn{1}+...\,,
\end{align}
with the coefficients $d_{0,k}$ given by
\begin{equation}\label{eq:d0k} 
d_{0,-1}=-2\,, \qquad  \widetilde{\mathcal{B}}[\mathcal{C}^{(0)}_{N\!P}](t)= -9(1+t)^2\tFo{-\tfrac{1}{2}}{\tfrac{5}{2}}{2}{-t(2+t)}=    \sum_{k=0}^\infty\frac{d_{0,k}}{k!}t^k  \,,
\end{equation}
 matching and extending the results of \cite{Dorigoni:2022cua} (modulo the trivial change in normalisation in footnote \ref{foot}) for the coefficients of the leading index non-perturbative terms, i.e.~all $\Dfn{s}$ terms in \eqref{eq:npSU} with $m=\ell$, presented in equation (3.26) of the same reference.

We can then construct the median resummation \eqref{median_resum_def} and express it via the transseries
\begin{equation}
    \mathcal{C}^{(0)}(N;\tau) = \PhiP{0}  + \sigma \,\PhiNP{0} \,,\label{eq:r0TS}
\end{equation}
where the associated transseries parameter $\sigma$ has value $\sigma = \pm i $ according to $\mbox{arg}(N){\gtrless}0$, precisely matching the transseries \eqref{eq:CrTS} found in \cite{Dorigoni:2022cua} by use of generating series methods.
Furthermore, the median resummation of said transseries can be written as the average of the two lateral resummations presented in \eqref{eq:Smed}, taking the form
\begin{align}
     \mathcal{C}^{(0)}(N;\tau)  &\notag =    \mathcal{S}_{med}[\mathcal{C}_P^{(0)}](N;\tau) = \intmed \NPfn{\sqrt{N}t}  \Big[-6t^2\tFo{-\tfrac{1}{2}}{\tfrac{5}{2}}{1}{t^2}\Big]\dd t\\
     &\label{eq:C0med}= \int_0^\infty \NPfn{\sqrt{N}t} \mbox{Re}\Big(-6t^2 \tFo{-\tfrac{1}{2}}{\tfrac{5}{2}}{1}{t^2}\Big)\dd t\,,
\end{align}
which will be retrieved from spectral methods later in section \ref{spectral_section}, see \eqref{eq:MellinM0}.

We can repeat this analysis for the contribution to \eqref{eq:originN1} coming from all ``sub-leading-index'' non-holomorphic Eisenstein series, i.e. all terms in \eqref{eq:originN1} with $m=\lfloor \ell/2\rfloor-1$ or equivalently consider \eqref{C_pert_expPhiR} with $r=1$:
\begin{align}
    \mathcal{C}^{(1)}_P(N,\tau) &\label{eq:genus1}= \sum_{k=0}^\infty b_{1,k}N^{-k-\frac{3}{2}} \eisen{\tfrac{3}{2}+k}\,,
\end{align}
where the coefficients $b_{1,k} = \tilde{b}_{k+2 ,\lfloor k/2\rfloor} $  have been computed in \cite{Dorigoni:2021guq} and are given by
\begin{equation}
    b_{1,k} \label{eq:b1k}\coloneqq -\frac{(k+1)^2(2k+13) \Gamma(k+\frac{5}{2})^2 }{2^{2k+6} \,3\pi^{\frac{3}{2}}\Gamma(k+3)}\,.
\end{equation}

The associated Borel transform \eqref{eq:BorelDef} is then given by
\begin{align}\label{Phi1_Borel}
   & \mathcal{B}[\mathcal{C}^{(1)}_P](t)= - \frac{1}{24\pi} \sum_{k=0}^\infty \frac{(k+1)(2k+13)\Gamma(k+\frac{5}{2})^2}{ \Gamma(k+1)\Gamma(k+3)}t^{2k+2} \\
    &\notag=-\frac{t^2}{8192}\Big[1248\,\tFo{\tfrac{5}{2}}{\tfrac{5}{2}}{3}{t^2}+3400\,t^2\,\tFo{\tfrac{7}{2}}{\tfrac{7}{2}}{4}{t^2}+1225\,t^4\,\tFo{\tfrac{9}{2}}{\tfrac{9}{2}}{5}{t^2}\Big]\,,
\end{align}
where again, due to the presence of these particular hypergeometric functions, we find the two Stokes directions $\mbox{arg}(t)=0$ and $\mbox{arg}(t)=\pi$ with corresponding logarithmic branch cuts starting at $t=\pm 1$. The singularity structure of $\mathcal{B}[\mathcal{C}^{(1)}_P](t)$ near $t=1$ is given by
\begin{align}
    \notag\mathcal{B}[\mathcal{C}_P^{(1)}](t) \sim &-\frac{1}{32\pi (1-t)^4}- \frac{3}{64\pi (1-t)^3}-\frac{(-77)}{512\pi (1-t)^2}-\frac{127}{1024\pi (1-t)}\\
    &+\widetilde{\mathcal{B}}[\mathcal{C}^{(1)}_{N\!P}](t-1) \frac{\log{(1-t)}}{\pi}+\text{reg}(t-1)\,,
\end{align}
where, following the discussion around \eqref{discont}, we have already interpreted the germ of analytic functions multiplying the logarithm as the Borel resummation of the non-perturbative sector given by
\begin{align}
& \widetilde{\mathcal{B}}[\mathcal{C}^{(1)}_{N\!P}](t)= \label{eq:NPC1}\\
&\notag - \frac{14 [60 t (t+2)+47] \, _2F_1\left(\frac{1}{2},\frac{1}{2};5;-t (t+2)\right)+[113 t (t+2)+269] \, _2F_1\left(\frac{1}{2},\frac{3}{2};5;-t (t+2)\right)}{8192 (t+1)^4}\,.
\end{align}

 We observe that, when compared to the Borel transform \eqref{Phi0_disc} for the case $r=0$, the order of the pole at $t=1$ has now increased to a fourth-order pole.
Similarly to \eqref{eq:NPC0}, from \eqref{eq:NPC1} we can now obtain the formal expansion of the sub-leading index non-perturbative sector
\begin{align}\label{Phi1_generating_series}
    \mathcal{C}^{(1)}_{N\!P}(N;\tau) &= \sum_{k=-3}^\infty d_{1,k} \,N^{-\frac{k+1}{2}}\Dfn{\tfrac{k+1}{2}} \,,
\end{align}
with coefficients
\begin{align}
&\label{eq:d1k} d_{1,-4}=\frac{1}{192}\,,\quad d_{1,-3}=\frac{3}{128}\,,\quad d_{1,-2}=-\frac{77}{512}\,,\quad d_{1,-1}=\frac{127}{1024}\,,\\
&\widetilde{\mathcal{B}}[\mathcal{C}^{(1)}_{N\!P}](t) =    \sum_{k=0}^\infty\frac{d_{1,k}}{k!}t^k  =-\frac{927}{8192} +\frac{3897}{16384}\, t-\frac{47217 }{65536} \,\frac{t^2}{2!}+O(t^3) \,.
\end{align}
 Once again these results extend those found in \cite{Dorigoni:2022cua} for the sub-leading diagonal of non-perturbative terms presented in equation (3.26) of that reference, i.e.~all $\Dfn{s}$ terms in \eqref{eq:npSU} with $m=\ell-1$.
 
In general, to reconstruct the non-perturbative completion of  $ \mathcal{C}^{(r)}_{P}(N;\tau)$, i.e. the non-perturbative sector completing the formal asymptotic expansion of ``$r$-subleading index'' non-holomorphic Eisenstein series \eqref{C_pert_expPhiR},
we start from the singular behaviour near $t=1$ of the corresponding Borel transform
\begin{align}
    \mathcal{B}[\mathcal{C}_P^{(r)}](t) \sim &-\frac{1}{\pi} \sum_{k=1}^{3r+1} \frac{d_{r,-k} (k-1)!}{ (1-t)^k}+\widetilde{\mathcal{B}}[\mathcal{C}^{(r)}_{N\!P}](t-1) \frac{\log{(1-t)}}{\pi}+\text{reg}(t-1)\,, \label{eq:DiscCr}
\end{align}
with
\begin{equation}
\widetilde{\mathcal{B}}[\mathcal{C}^{(r)}_{N\!P}](t) = \sum_{k=0}^\infty \frac{d_{r,k}}{k!} t^k\,.
\end{equation}
 Equation \eqref{eq:DiscCr} yields a difference in lateral Borel resummation of the form
 \begin{equation}
  (\mathcal{S}_+-\mathcal{S}_-)[\mathcal{C}^{(r)}_P](N;\tau)  = -2 i \mathcal{S}_0[\mathcal{C}^{(r)}_{N\!P}](N;\tau)\,.\label{eq:singR}
 \end{equation}
 The coefficients $d_{r,k}$ of the non-perturbative sector are then entirely encoded in the discontinuity equation \eqref{eq:singR}  of the Borel transform $ \mathcal{B}[\mathcal{C}_P^{(r)}](t)$ along  the Stokes line $t>0$,~i.e.
 \begin{align}\label{PhiR_generating_series}
   \mathcal{S}_0[\mathcal{C}^{(r)}_{N\!P}](N;\tau)\sim \mathcal{C}^{(r)}_{N\!P}(N;\tau) &= \sum_{k=-3r-1}^\infty d_{r,k}\,N^{-\frac{k+1}{2}}\Dfn{\tfrac{k+1}{2}} \,.
\end{align}

  Arguing as above, we must add to the lateral Borel resummation of the perturbative sector a suitable multiple of these non-perturbative terms to finally arrive at the modular invariant and unambiguous median transseries \eqref{eq:CrTS}
  \begin{align}
  &  \mathcal{C}^{(r)}(N;\tau) \notag= \PhiP{r}  + \sigma \,\PhiNP{r}  =    \mathcal{S}_{med}[\mathcal{C}_P^{(r)}](N;\tau) \\
   &\label{eq:CRmed} = \intmed \NPfn{\sqrt{N}t}\mathcal{B}[\mathcal{C}_P^{(r)}](t)\dd t   = \int_0^\infty \NPfn{\sqrt{N}t} \mbox{Re}\Big(\mathcal{B}[\mathcal{C}_P^{(r)}](t)\Big)\dd t\,,
  \end{align}
 where again the transseries parameter $\sigma=\pm i$ according to $\mbox{arg}(N)>0$ or $<0$.
 
\subsubsection*{Second integrated correlator}

We now apply the same resummation method to analyse a particular sector of the second integrated correlator $\mathcal{H}_N(\tau)$ presented in \eqref{eq:GHdef}.
The large-$N$ expansion of $\mathcal{H}_N(\tau)$ was initiated in \cite{Chester:2020vyz} and then conjectured in \cite{Alday:2023pet} to have the asymptotic perturbative form
\begin{equation}
\mathcal{H}_N(\tau) \sim 6N^2 +\mathcal{H}_N^h(\tau) + \mathcal{H}_N^i(\tau)\,,\label{eq:HlargeN}
\end{equation}
where the two different perturbative sectors $\mathcal{H}_N^h(\tau)$ and $\mathcal{H}_N^i(\tau) $ are formal modular invariant power series in respectively half-integer powers and integer powers in $1/N$.
In particular, the large-$N$ expansion of $\mathcal{H}_N^h(\tau)$ is of the same form  \eqref{C_pert_exp} for $\cC_{P} (N;\tau)$  and only contains non-holomorphic Eisenstein series, 
\begin{align}
     \mathcal{H}_{N}^{h} (\tau)  &\label{eq:H_pert_exp}=\sum_{r=0}^\infty N^{2-2r} \mathcal{H}^{(r)}_h (N;\tau)\,,\\
     \mathcal{H}^{(r)}_h (N;\tau) &\label{eq:H_pert_exp2} \coloneqq \sum_{k=0}^\infty a_{r,k} N^{-\frac{3}{2}-k} \eisen{\tfrac{3}{2}+k}\,.
\end{align}
For fixed value of $r$, the perturbative coefficients $a_{r,k}$ have been found in \cite{Alday:2023pet} exploiting an intriguing inhomogeneous Laplace difference equation relating this second integrated correlator $\mathcal{H}_N(\tau)$ to $\corr$.
In particular, for the leading-index non-holomorphic Eisenstein series we have
\begin{equation}
a_{0,k} \coloneqq {-}\frac{(k+1) (k+3) \Gamma \left(k-\frac{1}{2}\right) \Gamma \left(k+\frac{3}{2}\right)}{2^{2 k-3} \pi ^{3/2} \Gamma (k+1)}\,.
\end{equation}

Focusing for concreteness on the $r=0$ case, it is then straightforward to compute the corresponding Borel transform \eqref{eq:BorelDef},
\begin{equation}
\mathcal{B}[ \mathcal{H}_{h}^{(0)}](t) = 192\, t^2 \!\, _2F_1\left(-\tfrac{1}{2},\tfrac{3}{2};1\vert t^2\right)-48\, t^4\! \, _2F_1\left(\tfrac{1}{2},\tfrac{5}{2};2\vert t^2\right)\,\,
\end{equation}
from which we obtain the singular behaviour along the Stokes direction~$\mbox{arg}(t)=0$:
\begin{align}
    \mathcal{B}[\mathcal{H}_{h}^{(0)}](t) \sim &
   \label{eq:H0disc}-\frac{32}{\pi  (1-t)}+\widetilde{\mathcal{B}}[\mathcal{H}_{h,N\!P}^{(0)}](t-1) \frac{\log{(1-t)}}{\pi}+\text{reg}(t-1)\,,
\end{align}
and interpret the germ of analytic functions multiplying the logarithm as the Borel resummation of the non-perturbative sector given by
\begin{equation}
\widetilde{\mathcal{B}}[\mathcal{H}_{h,N\!P}^{(0)}](t) =\label{eq:H0NP} 48 (t+1)^2\! \left[4 \, _2F_1\left(-\tfrac{1}{2},\tfrac{3}{2};1\vert -t (t+2)\right){+}\, _2F_1\left(-\tfrac{1}{2},\tfrac{3}{2};2\vert-t (t+2)\right)\right].
\end{equation}

Following the same process as before, we extract from the singular behaviour \eqref{eq:H0disc} a novel formal series of non-perturbative corrections
\begin{align}\label{Phi0_generating_series}
\mathcal{H}_{h,N\!P}^{(0)}(N;\tau) &= \sum_{k=-1}^\infty h_{0,k}\,N^{-\frac{k+1}{2}} \Dfn{\tfrac{k+1}{2}} \\
    &\notag = 32\Dfn{0}+240N^{-\frac{1}{2}}\Dfn{\tfrac{1}{2}}+ 804N^{-1}\Dfn{1}+...\,,
\end{align}
with the coefficients $h_{0,k}$ given by
\begin{equation}\label{eq:d0k} 
h_{0,-1}=32\,, \qquad  \widetilde{\mathcal{B}}[\mathcal{H}_{h,N\!P}^{(0)}](t)=    \sum_{k=0}^\infty\frac{h_{0,k}}{k!}t^k =240+804 t + \frac{855}{2} \frac{t^2}{2!} +O(t^3)\,.
\end{equation}

A similar analysis can be carried out for higher values of $r>0$ to obtain the non-perturbative completion of the formal power series \eqref{eq:H_pert_exp}.
However, we stress that this process does not define the full non-perturbative completion for the second integrated correlator $\mathcal{H}_N(\tau)$. The procedure here discussed can only resum the formal power series in half-integer powers of $1/N$ contained in \eqref{eq:H_pert_exp}.
As discussed in \cite{Alday:2023pet}, besides the sector just discuss the large-$N$ expansion of the second integrated correlator \eqref{eq:HlargeN} contains the formal modular invariant power series $\mathcal{H}^{(i)}_N(\tau)$ in integer powers of $1/N$. 

Order by order in $1/N$, the coefficients
of $\mathcal{H}^{(i)}_N(\tau)$ are given by finite linear combinations of a different class of modular functions called \textit{generalised Eisenstein series}.
Generalised Eisenstein series have appeared in other string theory context, see e.g. \cite{Dorigoni:2019yoq,Dorigoni:2021jfr,Dorigoni:2022bcx}, and display a more complicated structure of perturbative and non-perturbative corrections \cite{Klinger-Logan:2018sjt,Dorigoni:2021ngn,Dorigoni:2023nhc,Fedosova:2023cab}.
In particular, we easily see that our resummation methods starting from the formal perturbative expansion \eqref{original_series} cannot be exploited to extract the non-perturbative completion to the sector $\mathcal{H}^{(i)}_N(\tau)$ of the second integrated correlator.

\subsection{Large-$N$ 't Hooft expansions} 
\label{tHooft_section}

In this section we consider the standard large-$N$ 't Hooft limit where $\lambda \coloneqq 4\pi N/\tau_2 = N g_{_{Y\!M}}^2$ is kept fixed as $N\to \infty$, starting from the transseries expansion \eqref{C_full_transseries}. We show that in this limit the non-perturbative resummation $\mathcal{C}^{(r)}(N;\tau)$ naturally encodes the strong coupling genus-$r$ 't Hooft transseries contribution, which includes an infinite tower of non-perturbative corrections of the form $e^{-2 \ell \sqrt{\lambda}}$ with $\ell \in \mathbb{N}$,  which can be interpreted as fundamental string world-sheet instantons. 
However, as a consequence of modularity we see that each $\mathcal{C}^{(r)}(N;\tau)$ also contains the strong coupling non-perturbative resummation of the ``dual'' genus-$r$ 't Hooft expansion, where the dual 't Hooft coupling $\tilde{\lambda} = (4\pi N)^2 / \lambda$ is kept fixed as $N\to \infty$. This resummation contains an infinite tower of non-perturbative corrections of the form $e^{-2 \ell \sqrt{\tilde{\lambda}}}$ with $\ell \in \mathbb{N}$,  which can be interpreted as an averaging of all dyonic-string world-sheet instantons. 

Since in the 't Hooft limit we keep fixed $\lambda = 4\pi N/\tau_2 = N g_{_{Y\!M}}^2$ as we send $N\to \infty$, we have that contributions from Yang-Mills instantons, of order $e^{- 8\pi^2 N |k| / \lambda}$ with instanton number $k\neq 0$, are exponentially suppressed. In the Fourier mode expansion of $\mathcal{C}^{(r)}(N;\tau)$ with respect to $\tau_1 = \theta/(2\pi)$, such contributions can be idenfitied with the $k^{th}$ Fourier mode. 
Hence in the 't Hooft limit we can restrict our attention to the analysis for the zero-mode sector of \eqref{eq:CRmed}, which is obtained from
\begin{equation}\label{eq:tHooftr}
\mathcal{I}^{(r)}(N;\lambda) \coloneqq \int_{-\frac{1}{2}}^{\frac{1}{2}}  \mathcal{C}^{(r)}(N;\tau) \,\dd \tau_1= \intmed  \Big[ \int_{-\frac{1}{2}}^{\frac{1}{2} }\NPfn{\sqrt{N}t}  \,\dd\tau_1 \Big]\mathcal{B}[\mathcal{C}^{(r)}_P ](t)\, \dd t\,.
\end{equation}

To extract the 't Hooft expansion of this expression, we need to compute the zero Fourier mode of the modular invariant modified Borel kernel $\NPfn{\sqrt{N}t}$. This calculation is presented in appendix \ref{sec:AppE} where we derive an explicit formula in \eqref{E0_formula}, here rewritten for convenience:
\begin{equation} 
    \int_{-\frac{1}{2}}^{\frac{1}{2} }\NPfn{\sqrt{N}t}  d\tau_1 = \frac{2}{e^{4t\sqrt{N\pi/\tau_2}}-1} + \mathcal{U}(\sqrt{N}t;\tau_2)\,.
\end{equation}
The function $\mathcal{U}(t;\tau_2)$ is given by either the contour integral representation \eqref{U_definition} or as an infinite sum over Bessel functions \eqref{eq:UBessel}-\eqref{eq:UBorel}. 
Substituting this expression for the zero-mode in \eqref{eq:tHooftr} we arrive at 
\begin{align}
\mathcal{I}^{(r)}(N;\lambda)\!&\phantom{:}=\label{eq:TrNP}I^{(r)}(\lambda)+N^{-1} \,\tilde{I}^{(r)}(\tilde{\lambda} ) \,,\\
I^{(r)}(\lambda) &\label{eq:INr} \coloneqq  \intmed \frac{2 }{e^{2t\sqrt{\lambda}}-1}\,\mathcal{B}[\mathcal{C}^{(r)}_P](t)\, \dd t \,,\\
\tilde{I}^{(r)}(\tilde{\lambda} )  & \label{eq:ItildeNr}\coloneqq \intmed \mathcal{U}\Big( t;\frac{\tilde{\lambda}}{4\pi}\Big)\,\mathcal{B}[\mathcal{C}^{(r)}_P](t)\,\dd t\,.
\end{align}
where we have introduced the dual 't Hooft coupling $\tilde{\lambda} = (4\pi N)^2 / \lambda$, and we have used the identity $\mathcal{U}(\sqrt{N}t;\frac{4\pi N}{\lambda})=N^{-1}\mathcal{U}(t;\frac{\tilde{\lambda}}{4\pi})$ which follows easily from \eqref{U_definition}.

If we plug this expression back into the full transseries representation \eqref{eq:NPdrk}, we obtain the complete 't Hooft limit expansion of the integrated correlator ${\corr}$:
\begin{align}
 &\label{eq:full0mode}\mathcal{I}(N;\lambda) \coloneqq \int_{-\frac{1}{2}}^{\frac{1}{2}}{\corr} \,\dd\tau_1  = \sum_{r=0}^\infty N^{2-2r} \mathcal{I}^{(r)}(N;\lambda) \\
&\notag =  \sum_{r=0}^\infty N^{2-2r} \intmed \frac{2 }{e^{2t\sqrt{\lambda}}-1}\,\mathcal{B}[\mathcal{C}^{(r)}_P](t)\, \dd t + \sum_{r=0}^\infty N^{1-2r}\intmed \mathcal{U}\Big( t;\frac{\tilde{\lambda}}{4\pi}\Big)\,\mathcal{B}[\mathcal{C}^{(r)}_P](t)\,\dd t\,.
\end{align}
We stress that thanks to our careful rewriting of the complete transseries \eqref{C_full_transseries}, the $N$-dependence has now been trivialised.
Furthermore, as we will shortly demonstrate, the modular invariant median resummation \eqref{eq:Smed} naturally leads to the median resummation of the genus-$r$ large 't Hooft coupling expansion, given by the contribution $I^{(r)}(\lambda)$ in \eqref{eq:TrNP}, plus the median resummation of the genus-$r$ dual 't Hooft coupling expansion,  encoded in the second term $\tilde{I}^{(r)}(\tilde{\lambda})$ in \eqref{eq:TrNP}. As already appreciated in \cite{Dorigoni:2022cua}, we note that modular invariance unifies in the single expression \eqref{eq:CRmed}, the seemingly different median 't Hooft-limit and dual 't Hooft-limit resummations studied in \cite{Dorigoni:2021guq} and \cite{Hatsuda:2022enx}.

\begin{figure}
    \centering
    \includegraphics[ scale = 0.24]{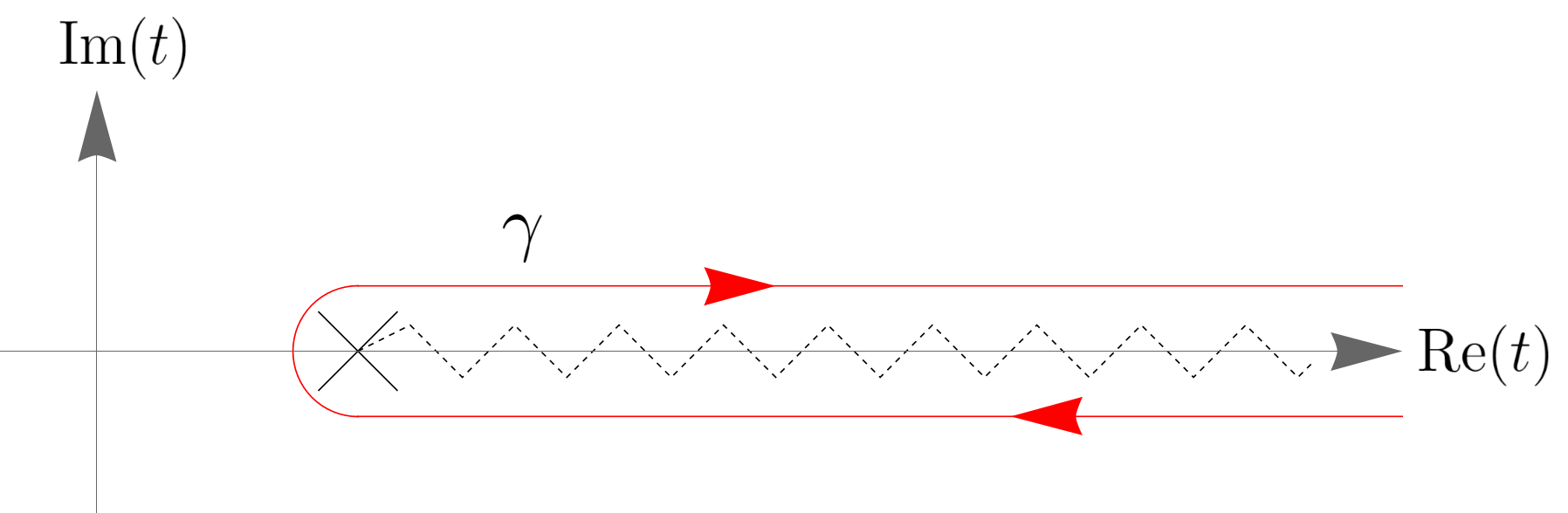}
    \caption{ Hankel contour $\gamma$ in the complex $t$-plane circling around the branch-cut singularity starting at $t=1$.}
    \label{fig:Hankel}
\end{figure}

To clarify these statements, let us separately consider the two terms in \eqref{eq:TrNP}.
Starting with the analysis for $I^{(r)}(\lambda) $, we first rewrite the median resummation making use of \eqref{median_resum_def},
\begin{equation}
I^{(r)}(\lambda) = \int_0^{\infty \pm i \epsilon} \frac{2 }{e^{2t \sqrt{\lambda}}-1}\,\mathcal{B}[\mathcal{C}^{(r)}_P](t)\, \dd t \, \mp \, \frac{1}{2}  \int_\gamma \frac{2 }{e^{2t\sqrt{\lambda}}-1}\,\mathcal{B}[\mathcal{C}^{(r)}_P](t)\, \dd t\,,\label{eq:INrsplit}
\end{equation}
where the Hankel contour $\gamma$ is given in figure \ref{fig:Hankel}.

The first term in this expression can be easily computed by expanding the Borel transform $\mathcal{B}[\mathcal{C}^{(r)}_P](t)$ using the definition \eqref{eq:BorelDef} and then integrating term by term using the identity
\begin{equation}\label{eq:moment1}
\int_0^\infty \frac{2 }{e^{2t \sqrt{\lambda}}-1}  t^k \dd t = 2^{-k } \lambda ^{-\frac{k }{2}-\frac{1}{2}} \Gamma (k +1) \zeta (k +1)\,,
\end{equation}
valid for $k\geq 0$.
In this way we arrive at the formal asymptotic power series expansion
\begin{align}
 \int_0^{\infty \pm i \epsilon} \frac{2 }{e^{2t \sqrt{\lambda}}-1}\,\mathcal{B}[\mathcal{C}^{(r)}_P](t)\, \dd t & \label{eq:pertThooft}
 \sim\sum_{k=0}^\infty b_{r,k} \, \xi(2k+3) \Big( \frac{\lambda}{4 \pi}\Big) ^{-k-\frac{3}{2}}\,,
\end{align} 
where $\xi(s)= \pi^{-s/2}\Gamma(s/2) \zeta(s) = \xi(1-s)$ is the completed zeta function as before.
For example, substituting the $r=0$ coefficients \eqref{eq:b0k} or the $r=1$ coefficients \eqref{eq:b1k}, one can check that \eqref{eq:pertThooft} reduces respectively to the standard genus-$0$ and genus-$1$ large 't Hooft coupling expansion of the integrated correlator presented in \cite{Dorigoni:2021guq}, cf. equations (5.26) and (5.28) of the reference. 

Alternatively, we note that
\begin{equation}
\frac{d}{d t} \left[  \frac{2 }{e^{2t \sqrt{\lambda}}-1} \right]= - \frac{4\sqrt{\lambda}}{4\sinh^2(t\sqrt{\lambda})}\,,
\end{equation}
hence we can integrate \eqref{eq:pertThooft} by parts\footnote{ It is easy to check from \eqref{Phi0_Borel} and \eqref{Phi1_Borel} that integration by parts does not produce any boundary contributions.} as 
\begin{equation}
 \int_0^{\infty \pm i \epsilon} \frac{2 }{e^{2t \sqrt{\lambda}}-1}\,\mathcal{B}[\mathcal{C}^{(r)}_P](t)\, \dd t = \sqrt{\lambda} \int_0^{\infty \pm i \epsilon} \frac{1}{4\sinh^2(t\sqrt{\lambda})}\Big( -4 \frac{d}{dt} \mathcal{B}[\mathcal{C}^{(r)}_P](t)\Big)\, \dd t \,.
\end{equation}
Again, substituting the Borel transform \eqref{Phi0_Borel} for $r=0$ or the $r=1$ counterpart \eqref{Phi1_Borel}, it is easy to see that this expression reduces precisely to the corresponding modified Borel resummation for the 't Hooft genus expansion considered in \cite{Dorigoni:2021guq}.

An important difference between the current analysis and the complete genus-$r$ large 't Hooft expansion is the absence in \eqref{eq:pertThooft} of finitely many positive powers of $\lambda$ which appear at any fixed genus. These powers will be retrieved by analysing the large-$\lambda$ perturbative expansion of the ``dual'' 't Hooft contribution \eqref{eq:ItildeNr}.

The second term in \eqref{eq:INrsplit} can again be computed from our general analysis, starting from the singular behaviour  \eqref{sing_t1} of the Borel transform $\mathcal{B}[\mathcal{C}^{(r)}_P](t)$, along the Stokes direction $t>0$.
To compute the contribution coming from the polar part of $\mathcal{B}[\mathcal{C}^{(r)}_P](t)$, we simply need the polylogarithm, $\polylog{k}{x}$, identity
\begin{equation}
\oint_{|t|=1}  \frac{2 }{e^{2 t \sqrt{\lambda}}-1} \frac{1}{(1-t)^k} \frac{\dd t}{{2\pi i}} = {-}2 \frac{ (4\lambda)^{\frac{k-1}{2}} \polylog{1-k}{e^{-2\sqrt{\lambda}}}}{\Gamma(k)}\,,\label{eq:tHooftPol}
\end{equation}
valid for $k\in \mathbb{N}$, while to evaluate the contribution coming from the discontinuity of the logarithmic singularity we first shift the contour of integration $t\to t+1$ and then use 
\begin{equation}
    \int_0^\infty \frac{2}{e^{2\sqrt{\lambda}(t+1)}-1}\, t^k\, \dd t = 2 (4\lambda)^{-\frac{k+1}{2}} k! \polylog{k+1}{e^{-2\sqrt{\lambda}}}\,,\label{eq:tHooftLog}
\end{equation}
valid for $\mbox{Re}(k)>0$.

We then express the second term in \eqref{eq:INrsplit} as
\begin{align}
&\mp \frac{1}{2} \int_\gamma \frac{2 }{e^{2t\sqrt{\lambda}}-1}\,\mathcal{B}[\mathcal{C}^{(r)}_P]({t})\, \dd t  = \pm  i \sum_{k=-3r-1}^{\infty} d_{r,k} (4\lambda)^{-\frac{k+1}{2}}  \polylog{k+1}{e^{-2\sqrt{\lambda}}}\,, 
\end{align}
which, as anticipated, encodes all non-perturbative contributions from worldsheet instantons.
It is easy to check that if we plug in the above expression the $r=0$ non-perturbative coefficients~\eqref{eq:d0k}, or similarly for the $r=1$  \eqref{eq:d1k}, we retrieve the necessary non-perturbative completions to the formal large-$\lambda$ perturbative expansion~\eqref{eq:pertThooft}, which had been obtained previously in~\cite{Dorigoni:2021guq,Hatsuda:2022enx} via resurgence analysis applied directly to~\eqref{eq:pertThooft}.

An analogous analysis can be carried out for the second contribution $\tilde{I}^{(r)}(\tilde\lambda)$. We again use the decomposition \eqref{median_resum_def} to rewrite equation  \eqref{eq:ItildeNr} as
\begin{equation}
\tilde{I}^{(r)}(\lambda) = \int_0^{\infty \pm i \epsilon} \mathcal{U}\Big( t;\frac{\tilde{\lambda}}{4\pi}\Big)\,\mathcal{B}[\mathcal{C}^{(r)}_P](t)\,\dd t \mp \frac{1}{2} \int_\gamma \mathcal{U}\Big( t;\frac{\tilde{\lambda}}{4\pi}\Big)\,\mathcal{B}[\mathcal{C}^{(r)}_P](t)\,\dd t \,,\label{eq:ItildeNrsplit}
\end{equation}
with the same Hankel contour $\gamma$ as shown in figure \ref{fig:Hankel}.

As before, the perturbative expansion at large-$\tilde{\lambda}$ is obtained from the first term  in the above equation.
Similarly to the analysis of \eqref{eq:pertThooft}, we expand the Borel transform $\mathcal{B}[\mathcal{C}^{(r)}_P](t)$  for $t$ small and integrate order by order using 
\begin{equation}
    \int_0^\infty  \mathcal{U}\Big( t;\frac{\tilde{\lambda}}{4\pi}\Big)\, t^{k} \,\dd t = \frac{ \Gamma\Big(\frac{k}{2}+1\Big)\Gamma\Big(\frac{k}{2}\Big)\zeta(k)}{4\pi} \tilde{\lambda}^{\frac{1-k}{2}}\,,
\end{equation}
valid for $\mbox{Re}(k)>{1}$ and proven in appendix \ref{sec:AppE}.
Given the above identity and the definition \eqref{eq:BorelDef} for the Borel transform, we arrive at the formal asymptotic power series expansion valid for large-$\tilde{\lambda}$:
\begin{equation}
\int_0^{\infty \pm i \epsilon} \mathcal{U}\Big( t;\frac{\tilde{\lambda}}{4\pi}\Big)\,\mathcal{B}[\mathcal{C}^{(r)}_P](t)\,\dd t \sim \sum_{k=0}^\infty b_{r,k} \,\xi(2k+2) \Big(\frac{ \tilde{\lambda} }{4 \pi}\Big)^{-k-\frac{1}{2}}\,.\label{eq:pertThooftD} 
\end{equation}
Using~\eqref{eq:UBorel}, we rewrite \eqref{eq:pertThooftD} as a Borel resummation of a modified Borel transform 
\begin{align}
\int_0^{\infty \pm i \epsilon} \mathcal{U}\Big( t;\frac{\tilde{\lambda}}{4\pi}\Big)\,\mathcal{B}[\mathcal{C}^{(r)}_P](t)\,\dd t &\label{eq:UtoBhat}= \sum_{n=1}^\infty \frac{\tilde{\lambda}}{\pi} \int_0^{\infty \pm i \epsilon} n \,e^{-2n x\sqrt{\tilde{\lambda}}} \,\widehat{\mathcal{B}}[\mathcal{C}^{(r)}_P](x)\,\dd x\,,\\
\widehat{\mathcal{B}}[\mathcal{C}^{(r)}_P](x) &\coloneqq \int_0^x  \frac{x}{t \sqrt{x^2-t^2}} \mathcal{B}[\mathcal{C}^{(r)}_P](t) \,\dd t\,. \label{eq:Bhat}
\end{align}

Substituting the $r=0$ coefficients \eqref{eq:b0k} or the $r=1$ coefficients \eqref{eq:b1k}, one can check that    \eqref{eq:pertThooftD} reduces respectively to the dual 't Hooft expansion at genus-$0$ or genus-$1$ of the integrated correlator presented in \cite{Collier:2022emf,Hatsuda:2022enx}. In particular, the integral representation \eqref{eq:UtoBhat} in terms of a modified Borel transform\footnote{
We note that the net effect of this modified Borel transform is to simply multiply the Taylor coefficients of the Borel transform by a ratio of gamma functions, i.e. $\int_0^x  \frac{x}{t \sqrt{x^2-t^2}}t^{2k+2} \,\dd t  = \frac{\sqrt{\pi }  \Gamma (k+1)}{2 \Gamma \left(k+\frac{3}{2}\right)} x^{2 k+2}$.} \eqref{eq:Bhat} of the original $ \mathcal{B}[\mathcal{C}^{(r)}_P](t)$ is identical (modulo some integration by parts) to the Borel integral-representation presented in \cite{Hatsuda:2022enx} for the dual 't Hooft genus expansion.

As in the previous decomposition \eqref{eq:INrsplit}, we find that second term in \eqref{eq:ItildeNrsplit} encodes non-perturbative effects at large $\tilde{\lambda}$ and can be computed from our general analysis of  the singular behaviour  \eqref{sing_t1} of the Borel transform $\mathcal{B}[\mathcal{C}^{(r)}_P](t)$, along the Stokes direction $t>0$.
The polar and logarithmic parts can be rewritten as
\begin{align}
\notag& \mp \frac{1}{2} \int_\gamma \mathcal{U}\Big( t;\frac{\tilde{\lambda}}{4\pi}\Big)\,\mathcal{B}[\mathcal{C}^{(r)}_P](t)\,\dd t \\
&\label{eq:UNP} = \mp i \sum_{k=1}^{3r+1} d_{r,-k}(k-1)! \oint_{|t|=1}  \mathcal{U}\Big( t;\frac{\tilde{\lambda}}{4\pi}\Big)\frac{ 1}{(1-t)^k} \frac{\dd t}{{2\pi i}}\pm i  \sum_{k=0}^\infty \frac{d_{r,k}}{k!} \int_0^\infty \mathcal{U}\Big( t+1;\frac{\tilde{\lambda}}{4\pi}\Big)\,t^k \,\dd t\,,
\end{align} 
with $\gamma$ the same Hankel contour in figure \ref{fig:Hankel}.
While it is easy to compute the polar part directly from \eqref{eq:UBessel} or \eqref{eq:UBorel}, we have not been able to find closed-form expressions akin to the 't Hooft limit analogues \eqref{eq:tHooftPol} and \eqref{eq:tHooftLog}. 

However, it is a straightforward task to plug in the above expression the genus-$0$ non-perturbative coefficients \eqref{eq:d0k}, or similarly for genus-$1$  \eqref{eq:d1k}, and upon expanding at large-$\tilde{\lambda}$ retrieve the necessary non-perturbative completions to the formal  perturbative expansion \eqref{eq:pertThooft}, obtained in \cite{Collier:2022emf,Hatsuda:2022enx} via resurgence analysis applied directly to \eqref{eq:pertThooftD}.
This is expected given that \eqref{eq:UtoBhat} relates the integral transform with kernel $ \mathcal{U}(t;x)$ to a more standard Borel resummation for the modified Borel transform $\widehat{\mathcal{B}}[\mathcal{C}^{(r)}_P](x) $ given in  \eqref{eq:Bhat}. The non-perturbative terms in the dual 't Hooft genus expansion which are encoded in \eqref{eq:UNP} can then be recovered directly from the directional Borel resummation \eqref{eq:UtoBhat}, precisely as discussed in \cite{Hatsuda:2022enx}.

As previously stated, our analysis shows that the modular invariant median resummation \eqref{eq:CRmed} of the ``r-subleading''-index non-holomorphic Eisenstein series \eqref{C_pert_expPhiR}, nicely encodes the median resummation of the genus-$r$ large 't Hooft expansion and dual large 't Hooft expansion.

As a last comment, we notice that when we substitute the dual 't Hooft perturbative expansion \eqref{eq:pertThooftD}  in the complete correlator $\mathcal{I}(\lambda)$, given in \eqref{eq:full0mode}, this can rewritten as a series in positive powers of $\lambda$ of the form:
\begin{align}
 \!\!\! \sum_{r'=0}^\infty  N^{1-2r'} \!\sum_{k=0}^\infty b_{r',k} \,\xi(2k+2) \Big(\frac{ \tilde{\lambda} }{4 \pi}\Big)^{\!-k-\frac{1}{2}} \!= \sum_{r=0}^\infty N^{2-2r} \!\sum_{k=0}^{r-1} b_{r-k-1,k} \,\xi (2 k+2) \,\Big(\frac{\lambda}{4\pi}\Big)^{\!k+\frac{1}{2}}\,,
\end{align}
where we changed the summation variable $r' =  r-k-1$, thus retrieving the ``missing'' powers of $\lambda$ from the complete perturbative genus-$r$ contribution in the 't Hooft limit, i.e. we have recovered the known perturbative genus expansion:
\begin{align}
\mathcal{I}(N;\lambda) &\sim \sum_{r=0}^\infty N^{2-2r} \mathcal{T}_P^{(r)}(\lambda)\,,\\
\mathcal{T}_P^{(r)}(\lambda) &\label{eq:tHooftR}\coloneqq \sum_{k=0}^{r-1} b_{r-k-1,k} \,\xi (2 k+2) \,\Big(\frac{\lambda}{4\pi}\Big)^{\!k+\frac{1}{2}}+\sum_{k=0}^\infty b_{r,k} \, \xi(2k+3) \Big( \frac{\lambda}{4 \pi}\Big) ^{-k-\frac{3}{2}}\,.
\end{align}
This particular combination of positive and negative powers of~$\lambda$ is a direct consequence of having rearranged the perturbative large-$N$ expansion~\eqref{eq:originN1}, where each non-holomorphic Eisenstein series~\eqref{Eisen_integral_rep} at large $\tau_2$ contributes~$\eisen{s} \sim \xi(2s)\tau_2^s  +  \xi(2s-1)\, \tau_2^{1-s}$. 

If we extend the definition of $b_{r,k}$ for $r\in \mathbb{N}$ to negative values of $k$ as 
\begin{equation}\label{eq:bhat}
\hat{b}_{r,k} \coloneqq \left\lbrace \begin{matrix}
b_{r+k+1,-k-2}\,,\qquad k\in \mathbb{Z}\,, k \leq -2\,,\\
 \!\!\!\!\!\!\!\!\! 0\,,\qquad \qquad \quad\! k=-1\,,\\
\!\!\!\!\!\!\!\! \!\!\!\!\!b_{r,k} \,,\qquad\qquad k\in \mathbb{N}\,,
\end{matrix}
\right.
\end{equation} 
and make use of the functional equation $\xi(s) = \xi(1-s)$, we can rewrite \eqref{eq:tHooftR} in the uniform manner  
\begin{equation}
    \mathcal{T}_P^{(r)}(\lambda)=\sum_{k=-r-1}^\infty \hat{b}_{r,k}\, \xi(2k+3) \Big(\frac{\lambda}{4\pi}\Big)^{-k-\tfrac{3}{2}}\,.
\end{equation}
Note that the would-be singular term $\xi(1)$ does not appear in the above expression since it multiplies $\hat{b}_{r,-1}=0$.
Interestingly, the coefficients $\hat{b}_{r,k}$ are in fact identical to the continuation of $b_{r,k}$ to negative values of $k$. 

The first non-trivial example of this fact appears at genus $r=2$ for which the coefficients $b_{2,k}$ have been computed in \cite{Dorigoni:2021guq} and are given by
\begin{equation}
    b_{2,k}=\frac{(k+1)^2( 20 k^2+ 208 k +219 )\Gamma(k+\frac{5}{2})\Gamma(k+\frac{11}{2})}{2^{2k+12}\,45\pi^{\frac{3}{2}}\Gamma(k+4)}\,.\label{eq:b2k}
\end{equation}
Given the definition \eqref{eq:bhat} and the explicit genus-$0$ and genus-$1$ coefficients \eqref{eq:b0k}-\eqref{eq:b1k}
we obtain directly
\begin{equation} \hat{b}_{2,-1}=0\,,\qquad \hat{b}_{2,-2}=b_{1,0}=-\frac{39}{2048\sqrt{\pi}}\,,\qquad \hat{b}_{2,-3}=b_{0,1}=\frac{15}{32\sqrt{\pi}}\,,
\end{equation}
while all other $\hat{b}_{2,k}= b_{k+3,-k-2} = 0 $ for $k\leq -4$ since $b_{r,k}$ with $r<0$ vanishes. Surprisingly if we substitute in \eqref{eq:b2k} negative values of $k$ we find precisely these numbers, i.e. we have $\hat{b}_{2,k} = b_{2,k}$. We have confirmed that the equality $\hat{b}_{r,k} = b_{r,k}$ for all $k\in \mathbb{Z}$, in particular for $k<0$, seems to persist at higher genus $r\geq 3$, however we do not have a proof of this statement, nor we understand the reasons behind it.

Since the analysis for the large-$N$ 't Hooft expansion of the sector $\mathcal{H}_N^h(\tau)$ for the second integrated correlator is pretty much identical to the above discussion, we will not repeat it here.
However, we want to highlight that the same analytic continuation  to negative $k$ of the perturbative coefficients $a_{r,k}$ for the ``r-subleading index'' non-Holomorphic Eisenstein series sector $\mathcal{H}^{(r)}_h (N;\tau)$ still seems to hold. 
That is if we take the analytic expressions for the coefficients $a_{r,k}$, which following \cite{Alday:2023pet} can be computed recursively in $r$ from the $b_{r,k}$, and continue them to negative values of $k$ exactly as in \eqref{eq:bhat},~i.e. for all the values of $r$ studied we find that $a_{r,-1}=0$ and $a_{r,k} = a_{r+k+1,-k-2}$ for  $k\in \mathbb{Z}\,, k \leq -2$ where again $a_{r,k}=0$ for $r<0$.

\section{Transseries from spectral representation}
\label{spectral_section}

In this section we show that the transseries \eqref{Eisen_integr_rep} can also be neatly derived starting from the spectral representation  \eqref{C_spec_rep} given in terms of the spectral overlap, $M_N(s)$, presented in \eqref{new_expr_spect_rep}. Firstly we show that the large-$N$ expansion of \eqref{C_spec_rep} naturally leads to a Borel resummed version of \eqref{C_full_transseries} thus demonstrating that the series over $r\in \mathbb{N}$ of all sectors $N^{2-2r} \mathcal{C}^{(r)}(N;\tau)$ is indeed Borel summable as previously stated. From here we derive the spectral representation of $ \mathcal{C}^{(r)}(N;\tau)$ for arbitrary $r$, thus reinterpreting the perturbative piece  \eqref{C_pert_expPhiR} as the polar contributions to the spectral integral, while the non-perturbative terms \eqref{eq:npSU1} arise as contributions from infinity. In this way, we produce a beautiful spectral-integral representation for the full transseries \eqref{C_full_transseries}.

\subsection{Spectral representation at large-$N$}

We start by analysing the large-$N$ expansion of the spectral representation \eqref{C_spec_rep} for the integrated correlator $\corr$.
Given the functional identity $E^*(s;\tau) = E^*(1-s;\tau)$ of the non-holomorphic Eisenstein series \eqref{Eisen_integral_rep} and the symmetry $M_N(s) = M_N(1-s)$ of the spectral overlap \eqref{new_expr_spect_rep}, at the price of losing manifest symmetry under $s\leftrightarrow 1-s$ we simply write
\begin{align}
    \corr &\,=  \frac{N(N-1)}{4}+ \int_{\Re(s)=\frac{1}{2}} \widetilde{M}_N(s)E^*(s;\tau)\frac{\dd s}{2\pi i}\,, \label{eq:spec1}\\
    \widetilde{M}_N(s) &\coloneqq   \frac{2^{1-2s} (2 s-1) \Gamma \left(\frac{3}{2}-s\right)}{\sqrt{\pi } \,\Gamma (-s)}\int_0^1 x^{s-3}(1-x)^N \tFo{s-1}{s}{2s}{x}\dd x \,.\label{eq:Mtilde}
\end{align}
where $M_N(s) = \tfrac{1}{2}(\widetilde{M}_N(s) +\widetilde{M}_N(1-s))$.

By changing integration variable in \eqref{eq:Mtilde} to $x= 1- e^{-\mu}$ with $\mu\in [0,\infty)$,  the spectral overlap $\widetilde{M}_N(s)$ takes immediately the form of a standard Borel resummation in $N$,
\begin{align}
\widetilde{M}_N(s) &\,=  \frac{2^{1-2s} (2 s-1) \Gamma \left(\frac{3}{2}-s\right)}{\sqrt{\pi } \,\Gamma (-s)} \int_0^\infty e^{-\mu N} \mu^{s} B(s;\mu)\,\dd  \mu  \label{eq:SpecBorel}\,,\\
B(s;\mu) &\coloneqq \mu^{-3} e^{-\mu } \left(\frac{1-e^{-\mu }}{\mu }\right)^{s-3} \, _2F_1\left(s-1,s;2 s\vert 1-e^{-\mu }\right)\,, \label{eq:SpecBorelTr}
\end{align}
where we notice for future reference that $\mbox{arg}(\mu)=0$ is not a Stokes direction for $B(s;\mu)$ and that $B(s;\mu) = {-} B(s;-\mu)$ as a consequence of the hypergometric function identity:
\begin{equation}
_2F_1(a,b;c\vert x) = (1-x)^{-a}\,_2F_1(a,c-b;c\big\vert \frac{x}{x-1})\,.
\end{equation}
As one can see by direct calculation, the simple expression \eqref{eq:SpecBorel} solves identically the Laplace difference equation \eqref{eq:LapDiffSpec} after some integrations by parts. 

The Borel transform $B(s;\mu) $ has an expansion for small $\mu$ of the form
\begin{equation}
B(s;\mu)  = \mu^{-3} -\frac{s (s+5)}{24  (2 s+1)} \mu^{-1} +\frac{  (s+2) (s+3) \left(5 s^2+37 s-12\right)}{5760 (2 s+1) (2 s+3)} \mu +O(\mu^3)\,.\label{eq:Bmu}
\end{equation}
In particular, we notice the potentially singular behaviour of $\mu^{s-3}$ at the origin of the Borel $\mu$-plane in \eqref{eq:SpecBorel}. However, it is easy to check that the regularisation procedure discussed in \eqref{I_regularisation}, amounts to regarding the term $\mu^{s}$ in \eqref{eq:SpecBorel} as a regulator to perform the Borel integral term by term and only after integration taking $s$ to lie on critical strip~$\mbox{Re}(s) = 1/2$.

We then compute the $\mu$-integral in \eqref{eq:spec1} by expanding the Borel transform for small $\mu$ and then integrate term by term arriving at the formal power series expansion
\begin{equation}
\widetilde{M}_N(s) = \sum_{r=0}^{2-2r} N^{2-2r-s} \mathcal{M}^{(r)}(s)\,,\label{eq:SpecLargeN}
\end{equation}
where for the first two orders we have
\begin{align}
\mathcal{M}^{(0)}(s) &\label{eq:M0}= \frac{2^{-2 s} (2s-1)^2 \Gamma(s)\Gamma(s+1) }{\sqrt{\pi }  \Gamma \left(s+\frac{1}{2}\right) } \,  \frac{\tan (\pi  s)}{(s-1)(s-2)} \,,\\
\mathcal{M}^{(1)}(s) &\label{eq:M1}= -\frac{2^{-2 s-1} s (s+5) (2s-1)^2 \Gamma(s)\Gamma(s+1) }{24 \sqrt{\pi }\Gamma\left(s+\frac{3}{2}\right)} \, \tan (\pi  s) \,,
\end{align}
while for $r\geq 2$ we find the general form
\begin{equation}
\mathcal{M}^{(r )}(s)   =\label{eq:Mr} \frac{ 2^{-2 s-4r} (2s-1)^2 \Gamma (s+2r)\Gamma(s+1)   } {\sqrt{\pi }\, \Gamma \left(s+r+\frac{1}{2}\right)} P^{(r)}(s) \tan (\pi  s)  \,,
\end{equation}
for some polynomials $P^{(r)}(s)$ of degree $2r-2$. For example we find
\begin{align} 
P^{(2)}(s) &\notag = \frac{5 s^2+37 s-12}{90}\,, \qquad P^{(3)}(s) = -\frac{35 s^4+462 s^3+1153 s^2+750 s-720}{5670}\,,\\
P^{(4)}(s) & = \frac{175 s^6+3745 s^5+24579 s^4+71327 s^3+84086 s^2+12648 s-60480}{340200}\,.
\end{align}

A key observation about the functions $\mathcal{M}^{(r)}(s)$ is that they are all analytic functions in the strip $1/2 < \mbox{Re}(s) < 3/2$ and they all have a simple zero at $s=1$, apart from the case $r=0$ for which we have 
\begin{equation}
\lim_{s\to 1}\mathcal{M}^{(0)}(s)= -\frac{1}{2}\,.
\end{equation}
This means that if we substitute the large-$N$ expansion for the spectral overlap \eqref{eq:SpecLargeN} in the integral representation \eqref{eq:spec1}, we can push the contour of integration to $\mbox{Re}(s)=1+\epsilon$ with $0<\epsilon<\tfrac{1}{2}$. 

Only for the case $r=0$ we have to be careful, since by doing so we pick up the residue at $s=1$ coming from the non-zero limit $\lim_{s\to 1}\mathcal{M}^{(0)}(s)= -\frac{1}{2}$ multiplying the simple pole of the non-holomorphic Eisenstein series \eqref{Eisen_integral_rep}, which in our normalisation has residue 
\begin{equation}
\mbox{res}_{s=1}E^*(s;\tau) =\frac{1}{2}\,.
\end{equation}
This residue at $s=1$, coming from the case $r=0$, combines with the constant term $N(N-1)/4$ so that \eqref{eq:spec1} can be rewritten as
\begin{equation}
\corr =  \frac{N^2}{4}+ \sum_{r=0}^\infty N^{2-2r} \int_{\Re(s)=1+\epsilon}\mathcal{M}^{(r)}(s) N^{-s} E^*(s;\tau)\frac{\dd s}{2\pi i}\,,\label{eq:largeNSpec}
\end{equation}
with $0<\epsilon< \tfrac{1}{2}$.

Few comments are in order at this point:
\vspace{-0.2cm}
\begin{itemize}
\item[(i)] the expansion in even powers of $N$ (modulo the Mellin-like term $N^{-s}$) is a direct consequence of the previously noted fact that the Borel transform is an odd function of the Borel variable $\mu$;\vspace{-0.2cm}
\item[(ii)] by comparing the expansion \eqref{eq:largeNSpec} with \eqref{C_full_transseries} it appears manifest, and it is proven in the next section, that $\mathcal{M}^{(r)}(s) N^{-s}$ must be the spectral overlap  of $\mathcal{C}^{(r)}(N;\tau)$;\vspace{-0.2cm}
\item[(iii)]  once we identify \eqref{C_full_transseries} with the spectral representation for  \eqref{eq:largeNSpec}, we deduce, as previously stated,  that the sum over $r$ in \eqref{C_full_transseries} is Borel summable: the resummation over $r\in \mathbb{N}$ of the asymptotic series of spectral overlaps \eqref{eq:SpecLargeN}  is understood via the standard Borel transform \eqref{eq:SpecBorel}.
\end{itemize}

We now move to show that \eqref{eq:largeNSpec}  in fact provides for the spectral representation of the modular invariant transseries \eqref{C_full_transseries}.

\subsection{Large-$N$ transseries from a spectral perspective}
\label{sec:SpecRes}

By comparing the large-$N$ spectral representation \eqref{eq:largeNSpec} with the previously analysed modular invariant transseries representation \eqref{C_full_transseries}, we immediately see that the spectral representation for each $\mathcal{C}^{(r)}(N;\tau)$ must take the form
\begin{align}\label{strong_coup_large_N}
     \mathcal{C}^{(r)}(N;\tau)& = \int_{\Re(s) = 1+\epsilon} \mathcal{M}^{(r)}(s)N^{-s}\eisen{s}\frac{\dd s}{2\pi i}\,,
\end{align}
where the spectral overlaps,~$\mathcal{M}^{(r)}(s)$,  are precisely the ones we obtained from the large-$N$ expansion~\eqref{eq:SpecLargeN}.

To prove that \eqref{strong_coup_large_N} is in fact correct, we relate this expression directly with the median resummation formula \eqref{eq:CRmed}, for example we show that when $r=0$ the expression \eqref{strong_coup_large_N} reduces identically to \eqref{eq:C0med}. 
To this end we make use of the integral representation \eqref{Eisen_integr_rep} for $N^{-s} E^*(s;\tau)$, presently appropriate since $\Re(s) = 1+\epsilon$, and we rewrite \eqref{strong_coup_large_N} as to obtain
\begin{equation}
     \mathcal{C}^{(r)}(N;\tau)= \int_0^\infty   \NPfn{\sqrt{N}t} \Big[ \int_{\Re(s) = 1+\epsilon}  (4t)^{2s-1} \frac{2\Gamma(s) \mathcal{M}^{(r)}(s)}{\Gamma(2s)}\frac{\dd s}{2\pi i}\Big]\,\dd t\,.\label{eq:Crspec}
\end{equation}
By comparing this expression with the median resummation formula \eqref{eq:CRmed}, we conclude that for $t>0$ we must have 
\begin{equation}
\mbox{Re}\Big( \mathcal{B}[\mathcal{C}^{(r)}_P](t) \Big) =  \int_{\Re(s) = 1+\epsilon}  (4t)^{2s-1} \frac{2\Gamma(s)\mathcal{M}^{(r)}(s)}{\Gamma(2s)}\frac{\dd s}{2\pi i} \,,\label{eq:MedianSpec}
\end{equation}
i.e. the Borel transform for the median resummation is related to the inverse Mellin transform of the spectral overlap $\mathcal{M}^{(r)}(s)$.

For concreteness, let us consider the explicit cases $r=0$ presented in \eqref{eq:M0}, for $r>0$ the story is identical with just slightly different expressions \eqref{eq:M1}-\eqref{eq:Mr}.  Starting from \eqref{eq:M0}, we need to compute:
\begin{align}
 \int_{\Re(s) = 1+\epsilon}  \!\!\!\! (4t)^{2s-1} \frac{2\Gamma(s)\mathcal{M}^{(0)}(s)}{\Gamma(2s)}\frac{\dd s}{2\pi i} &\label{eq:B0Re}=   \int_{\Re(s) = 1+\epsilon} \!\!\!\! t^{2 s-1}  \frac{4 \Gamma (s-2) \Gamma (s+1)}{\Gamma \left(s-\frac{1}{2}\right)^2} \tan (\pi  s)  \frac{\dd s}{2\pi i} \,.
\end{align}

These integrals can be understood as inverse Mellin transforms in the $s$ variable and can be evaluated by closing the contour of integration in a suitable manner. For $0<t<1$ we must close the contour of integration to the right half-plane $\mbox{Re}(s)>1$, hence picking up the simple poles coming from $\tan (\pi  s) $ and located at $s = n+1/2$ with $n\in \mathbb{N}^{>0}$. Similarly, for $t>1$ the contour of integration must be closed in the left half-plane $\mbox{Re}(s)<0$. The poles from $\tan (\pi  s) $  are now compensated by the gamma functions at denominator, and we are left with the simple poles located at $s=-n$ with $n\in \mathbb{N}^{>0}$ coming from the double poles of the gamma functions at numerator combined with the simple zeroes of $\tan(\pi s)$. 
In both cases it is possible to show that the contribution at infinity vanishes.

Proceeding as just described, we see that \eqref{eq:B0Re} is given by
\begin{align}\label{eq:MellinM0}
   \int_{\Re(s) = 1+\epsilon}  \!\!\!\! (4t)^{2s-1} \frac{2\Gamma(s)\mathcal{M}^{(0)}(s)}{\Gamma(2s)}\frac{\dd s}{2\pi i} &=\begin{cases}
			-6 t^2 \, _2F_1\left(-\frac{1}{2},\frac{5}{2};1\vert \,t^2\right)\,, & \qquad  0<t<1 \,,\\
         -\frac{3}{8 t^3} \,_2F_1\left(\frac{5}{2},\frac{5}{2};4\,\vert \,t^{-2}\right)\,, &\qquad  t\geq 1\,.
		 \end{cases}
\end{align}
Since we want to show that the spectral representation \eqref{eq:Crspec} coincides with the median transseries resummation, then for $r=0$ we must have that \eqref{eq:MellinM0} equals \eqref{eq:C0med}.
In particular, we see that \eqref{eq:MellinM0} is identical to the Borel transform $\mathcal{B}[\mathcal{C}^{(0)}](t)$ given in \eqref{Phi0_Borel} for $0<t<1$.
Furthermore, as already discussed in detail we know that $\mathcal{B}[\mathcal{C}^{(0)}](t)$ has a branch-cut singularity starting at $t=1$ with a discontinuity \eqref{Phi0_disc} which is purely imaginary for $t>1$. From the integral representation for the hypergeometric function we can then derive that for $t\in[0,\infty)$ we have
\begin{align}
&\notag \mbox{Re}\Big(\mathcal{B}[\mathcal{C}^{(0)}](t)\Big) =  
 \frac{1}{2} (-6t^2)  \lim_{\epsilon\to 0^+} \Big[\,_2F_1\left(-\frac{1}{2},\frac{5}{2};1\vert (t+i\epsilon)^2\right) +\!\,_2F_1\left(-\frac{1}{2},\frac{5}{2};1\vert (t-i\epsilon)^2\right)  \Big]\\
 & = \left\lbrace 
 \begin{matrix} 
  -6 t^2 \, _2F_1\left(-\frac{1}{2},\frac{5}{2};1\vert \,t^2\right)\,, &  \qquad  \quad \quad \! \!0<t<1 \,,\\ 
\phantom{\Big[} -\dfrac{3}{8 t^3} \,_2F_1\left(\frac{5}{2},\frac{5}{2};4\,\vert \, t^{-2}\right)\,, &\qquad  t\geq 1\,,
\end{matrix}\right.
\end{align}
hence we conclude that indeed \eqref{eq:MellinM0} is identical to $ \mbox{Re}\Big(\mathcal{B}[\mathcal{C}^{(0)}](t)\Big)$. For higher values of $r>0$, we have checked that indeed the spectral representation \eqref{strong_coup_large_N} coincides precisely with the modular invariant median resummation \eqref{eq:CRmed} for $\mathcal{C}^{(r)}(N;\tau)$.

Before concluding, we also want  to clarify how to extract directly from the spectral representation \eqref{strong_coup_large_N} the formal transseries expansion \eqref{eq:CrTS} in terms of the perturbative, non-holomorphic Eisenstein series part \eqref{C_pert_expPhiR} and the non-perturbative sector \eqref{eq:NPdrk}. 
We again focus concretely on the case $r=0$, although everything we say can be applied to arbitrary $r$.
Substituting the expression \eqref{eq:M0} for  $\mathcal{M}^{(0)}(s)$ in \eqref{strong_coup_large_N} and massaging some of the gamma functions we arrive at
\begin{equation}\label{eq:C0Spec}
     \mathcal{C}^{(0)}(N;\tau)\,{=}\!  \int_{\Re(s) = 1+\epsilon}\!\!\!\!\!  \frac{2^{2-4 s} \Gamma (2 s)}{\Gamma (s)} \,\frac{2 \Gamma (s-2) \Gamma (s+1)}{\Gamma \left(s-\frac{1}{2}\right)^2}\,  \tan (\pi  s)N^{-s}\eisen{s}\frac{\dd s}{2\pi i}\,.
\end{equation}

The perturbative part \eqref{eq:PhiP0} is clearly obtained by formally closing the contour of integration to the right half-plane, $\mbox{Re}(s)>1$, and summing over minus (due to the orientation of the integration contour) the residues from the poles coming from $\tan (\pi  s)$ and located at $s= k+3/2$ with $k\in \mathbb{N}$. A simple residue calculation immediately shows  
\begin{align}
{-}\mbox{res}_{s= k+ \frac{3}{2}}\Big[  \frac{2^{2-4 s} \Gamma (2 s)}{\Gamma (s)} \,\frac{2 \Gamma (s-2) \Gamma (s+1)}{\Gamma \left(s-\frac{1}{2}\right)^2}  \tan (\pi  s)N^{-s}\eisen{s}\!\Big] \!=\! b_{0,k}N^{-\frac{3}{2}-k} \eisen{\tfrac{3}{2}+k},
\end{align}
for $k\in \mathbb{N}$ and where the coefficients $b_{0,k}$ are exactly the ones given in \eqref{eq:b0k}.

While the formal sum over all residues on the positive real $s$-axis reproduces the perturbative series $\mathcal{C}^{(0)}_P(N;\tau)$ in \eqref{eq:PhiP0}, the non-perturbative sector $\mathcal{C}^{(0)}_{N\!P}(N;\tau)$ in \eqref{Phi0_generating_series} is encoded in the formal contribution at infinity.
This can be made explicit by first substituting in \eqref{eq:C0Spec} the lattice sum representation \eqref{Eisen_integral_rep} for the non-holomorphic Eisenstein series $\eisen{s}$, and then evaluating the $s$-integral via a saddle-point analysis.
One can easily see that at large-$N$ the spectral representation \eqref{eq:C0Spec} behaves as
\begin{align}
&\notag\frac{2^{2-4 s}  \Gamma (2 s)\Gamma (s-2) \Gamma (s+1)}{\Gamma \left(s-\frac{1}{2}\right)^2}\,  (N Y_{mn}(\tau))^{-s} \\
&\label{eq:saddleExp} \sim\exp\Big[ 2s \big(\log s- \log ( 2 \sqrt{N Y_{mn}(\tau)}  ) -1\big) \Big] \frac{ 4\sqrt{\pi} }{\sqrt{s}}\Big( 1+ \frac{55}{24 s} +O(s^{-2})\Big) \,.
\end{align}
In the limit $N\gg1$, the argument of the exponential function has a saddle point located at $s=s_\star$ with $|s_\star|\gg 1$:
\begin{equation}
s_\star \coloneqq 2 \sqrt{N Y_{mn}(\tau)}\,.\label{eq:saddle}
\end{equation}
By evaluating \eqref{eq:saddleExp}  at the saddle point $s=s_\star$, we find that the leading growth is given by $\exp( -4 \sqrt{ N Y_{mn}(\tau)})$, i.e. precisely the exponential suppression factor of the non-perturbative~$D_N(s;\tau)$ modular functions~\eqref{Dfn_def}.

We notice furthermore, that the  term $ \tan (\pi  s)$ in \eqref{eq:C0Spec} is the realisation of the transseries parameter $\sigma$ in \eqref{eq:r0TS}. If we evaluate this factor at the saddle location \eqref{eq:saddle}, we see that it reduces to
\begin{equation}
\tan (\pi  s_\star) \stackrel{|N|\gg1 }{\longrightarrow} \left\lbrace 
\begin{matrix}
+i \,, \qquad\qquad \text{ arg}(N)>0\,,\\
-i \,, \qquad\qquad \text{ arg}(N)<0\,.
\end{matrix}
\right.
\end{equation}

It is straightforward to expand around the saddle point by changing integration variables to 
\begin{equation}
s=s_\star+ i \,(N Y_{mn})^{\frac{1}{4}} \delta s\,,
\end{equation}
 so that the integral representation \eqref{eq:C0Spec} reduces at large-$N$ to a gaussian integral in the fluctuations $\delta s$ timed by an infinite formal series of perturbative corrections, which can be evaluated to arbitrary high order in $(N Y_{mn})^{-1}$ thus recovering the non-perturbative sector $ \mathcal{C}^{(0)}_{N\!P}(N;\tau)$ previously obtained in \eqref{Phi0_generating_series} via resurgent analysis arguments.

We conclude with a quicker and suggestive, albeit not completely rigorous way of showing that the non-perturbative coefficients $d_{r,k}$ of the non-perturbative sector \eqref{PhiR_generating_series} are encoded directly at the level of spectral overlap $\mathcal{M}^{(r)}(s)$.
Focusing again on the showcase example where $r=0$, we analyse the integrand of the spectral decomposition \eqref{eq:C0Spec} and we expand it at large-$s$ in the following manner,
\begin{align}
 \frac{2^{2-4 s} \Gamma (2 s)}{\Gamma (s)} \,\frac{2 \Gamma (s-2) \Gamma (s+1)}{\Gamma \left(s-\frac{1}{2}\right)^2}=
  \frac{2^{2-4 s} \Gamma (2 s)}{\Gamma (s)} \left( \sum_{\ell=0}^\infty \delta_\ell \,(2s)^{-\ell } \right)\,,\label{eq:largeSspec0}
 \end{align}
 where the first few $\delta_\ell$ coefficients are given by
 \begin{equation}
 \delta_0 =2\,, \qquad \delta_1 = 9\,,\qquad \delta_2 = \frac{153}{4}\,.
 \end{equation}
 
 We find that the non-perturbative coefficients $d_{0,k}$ presented in \eqref{Phi0_generating_series} are in fact encoded entirely in the above expression via
 \begin{equation}
 d_{0,k} = \sum_{\ell=0}^{k+1} S_{k+1}^{(\ell)} \, \delta_\ell\,,\label{eq:Stir}
 \end{equation}
 where $S_k^{(\ell)}$ denotes the Stirling number of the first kind.
 This identity follows from the properties of the Stirling numbers: the particular linear combination of the coefficients $\delta_\ell$  defined in \eqref{eq:Stir} allows us to rewrite \eqref{eq:largeSspec0} in the alternative large-$s$ expansion
 \begin{align}
 \frac{2^{2-4 s} \Gamma (2 s)}{\Gamma (s)}\left( \sum_{\ell=0}^\infty \delta_\ell \,(2s)^{-\ell }\right) &\notag = \frac{2^{2-4 s} \Gamma (2 s)}{\Gamma (s)}  \left(\, \sum_{k=-1}^\infty d_{0,k} \prod_{i=1}^{k+1} (2s-i)^{-1}\right)\\
 &= \sum_{k=-1}^\infty d_{0,k} \frac{2^{2-4 s} \Gamma (2 s - k-1)}{\Gamma (s)}\,.\label{eq:SpecOverExp}
 \end{align}

As a last step, we notice that the particular factor in the summand, 
$$ \frac{2^{2-4 s} \Gamma (2 s - k-1) }{\Gamma (s)}\,,$$
 is precisely the spectral overlap with the non-holomorphic Eisenstein series of the modular invariant function $N^{-\frac{k+1}{2}} D_N(\frac{k+1}{2};\tau)$ computed in \cite{Paul:2023rka}, for general index given by
\begin{equation}\label{eq:SpecDN}
D_N(p;\tau) = \langle D_N(p)\rangle + \int_{\Re(s) = \frac{1}{2}+\epsilon}  \frac{2^{2-4 s} \Gamma (2 s - 2p)}{\Gamma (s)}N^{p-s} \eisen{s} \frac{\dd s}{2\pi i}\,.
\end{equation}
A similar analysis can be carried out starting from $\mathcal{M}^{(r)}(s)$ with $r\geq1 $ to retrieve the non-perturbative coefficients $d_{r,k}$. 

While the expansion \eqref{eq:SpecOverExp} is rather suggestive, we stress that this result does not quite show how to rigorously obtain the non-perturbative sector \eqref{PhiR_generating_series} from the spectral representation \eqref{eq:C0Spec}, unlike the previous two arguments exploiting either median resummation \eqref{eq:MedianSpec} or saddle point analysis. In fact, we see from \eqref{eq:SpecDN} that the spectral representation for $D_N(p;\tau)$ is given by an integral over the line $\mbox{Re}(s)=1/2$, while here we have obtained indeed the correct spectral overlaps in the expansion \eqref{eq:SpecOverExp}, but only in the limit $\mbox{Re}(s)\gg 1$ and without any apparent trace of the spectral average $\langle D_N(p)\rangle$.

We find it rather beautiful how the simple spectral representation \eqref{strong_coup_large_N} encodes in these various interesting ways the perturbative non-holomorphic Eisenstein series part \eqref{C_pert_expPhiR}, the non-perturbative $D_N(p;\tau)$ sector \eqref{eq:NPdrk} and, as a matter of fact, the complete median resummation transseries representation \eqref{eq:CrTS}.

\section{Conclusions}
\label{sec:Conc}
This paper contains two main results. 
Firstly, we have defined a modified modular invariant Borel kernel \eqref{resummation_def}, thanks to which we have been able to reconstruct the full non-perturbative large-$N$ transseries representation of the first integrated correlator $\corr$  starting from its formal perturbative expansion defined in \eqref{eq:GHdef}. With the same method we resummed the non-holomorphic Eisenstein series sector of the second integrated correlator $\mathcal{H}_N(\tau)$.
As a consequence of our modular invariant approach, both  the resurgent genus expansion transseries at large  't Hooft-coupling $\lambda$ and large dual-'t Hooft-coupling $\tilde{\lambda}$ are neatly encoded in our modular invariant non-perturbative resummations.

Secondly, we analysed the problem of extracting the large-$N$ transseries expansion of the first integrated correlator $\corr$ starting from its spectral decomposition in terms of $L^2$-normalisable eigenfunctions of the hyperbolic Laplace operator $\Delta_\tau$. We have rewritten the spectral overlap function as a simple Borel transform in the number of colours $N$, from which we have extracted the large-$N$ spectral overlaps for the transseries we previously constructed.
Both the perturbative and non-perturbative formal yet modular invariant large-$N$ expansions are nicely encoded in these spectral overlaps.

A first interesting follow-up to our work, is the analysis of the first integrated correlator $\corr$ for other classical gauge groups, i.e. for $G=SO(N)$ and $USp(2N)$. Generating series over $N$ have been constructed for all these cases in \cite{Dorigoni:2022cua}, relying heavily on \cite{Alday:2021vfb}. 
Similarly to the $SU(N)$ case, starting from the generating series it is possible to construct the non-perturbative completions at large-$N$ for all classical gauge groups. Although the non-perturbative sectors for any classical gauge groups all live in the same class of exponentially suppressed modular functions,~$D_N(s;\tau)$, we find many subtle differences between the complete transseries expansions for the various gauge groups, for example in the transseries parameters. We believe that thanks to our modular invariant resurgent analysis approach, it should be possible to explain why and how the non-perturbative sectors vary amongst the classical gauge groups, $G=SU(N),SO(N)$ and $USp(2N)$. 

Secondly, we believe that our resummation method can help with the study of other integrated correlators.
In particular, exact expressions have been found \cite{Paul:2022piq,Brown:2023cpz,Paul:2023rka,Brown:2023why} for certain integrated four-point functions of two superconformal primary operators, $\mathcal{O}_2$, and two identical higher conformal dimension operators denoted by $\mathcal{O}_p$, with $p\in \mathbb{N}$ and $p \geq 2$. 
An interesting story \cite{Paul:2023rka,Brown:2023why} arises in the large charge limit, $p\gg1$, at fixed $N$ of these observables: for $N$ odd the modular invariant large-$p$ expansion takes a form very similar to \eqref{eq:originN1} (where the r\^ole of $N$ in that equation is now played by $p$), while for $N$ even the modular invariant perturbative expansion in $1/p$ does become convergent.

Independently from the nature of the large-$p$ perturbative expansion, \cite{Paul:2023rka,Brown:2023why} showed that  the large-charge expansion of these integrated correlators does in fact contain non-perturbative corrections of the form $D_{p/2}(r;\tau)$. This phenomenon is very likely a nice example of ``Cheshire cat'' resurgence \cite{Dunne:2016jsr,Kozcaz:2016wvy,Dorigoni:2017smz,Dorigoni:2019kux,Fujimori:2022qij}, where a would-be asymptotic perturbative tail disappears at special values of some external parameters, in this case for $N$ an even integer, while the body of the resurgent structure lingers on. We believe that the modular invariant resummation procedure here introduced can be exploited to resum the large-$p$ expansion of these higher-charge integrated correlators and manifest the Cheshire cat nature in $N$ of the resurgent perturbative large-$p$ expansion.

Another important open problem is understanding the complete large-$N$ expansion for the second integrated correlator $\mathcal{H}_N(\tau)$ in \eqref{eq:GHdef}.
Focusing at first on the sector $\mathcal{H}_N^{h}(\tau)$, containing only half-integer powers in $1/N$ and non-holomorphic Eisenstein series, we believe that exploiting the Laplace difference equation conjectured in \cite{Alday:2023pet} it should be possible to obtain a Borel-resummed spectral representation, hopefully akin to the first integrated correlator  \eqref{eq:SpecBorel}. Although $\mathcal{H}_N^{h}(\tau)$ is only part of the full integrated correlator, this would be a big step towards understanding the $N$ dependence of $\mathcal{H}_N(\tau)$.

However, a tougher hurdle is posed by the sector $\mathcal{H}_N^{i}(\tau)$, which contains integer powers in $1/N$ whose coefficients are generalised Eisenstein series. While in \cite{Alday:2023pet} it was conjectured that $\mathcal{H}_N^{i}(\tau)$ admits a lattice sum representation order by order in $1/N$, the systematic is still not fully understood and no useful spectral decomposition is available.
Furthermore \cite{Alday:2023pet} proposed an intriguing equation which relates $\Delta_\tau \mathcal{H}_N^{i}(\tau)$ to $\corr^2$.
From the transseries representation \eqref{eq:TSgen}, we then expect a novel class of modular invariant non-perturbative corrections\footnote{We thank Congkao Wen and Michael B. Green for related discussions.} to appear in the resummation of $\mathcal{H}_N^{i}(\tau)$ of the form
\begin{equation}
(\Delta_\tau - \lambda) F(\tau) = D_{t_1}(s_1;\tau) D_{t_2}(s_2;\tau) \,.
\end{equation}
Here $t_1,t_2$  are auxiliary parameters which can be set either to $N$ as to retrieve a non-perturbative term $D_N(s;\tau)$ or to $0$ as to reduce from \eqref{Dfn_def} to a standard non-holomorphic Eisenstein series.
This analysis is well beyond the scope of this paper and most definitely deserves further investigations.


\medskip
\subsection*{Acknowledgements}
We thank Congkao Wen, Hynek Paul and Michael B. Green for useful conversations and to Congkao Wen for comments on the draft.
We are particularly grateful to the Galileo Galilei Institute for Theoretical Physics for the hospitality and the INFN for partial support during the completion of this work, as well as to all the participants of the GGI programme ``Resurgence and Modularity in QFT and String Theory'' for stimulating discussions.

\appendix

\section{An alternative spectral overlap}
 \label{spectral_rep_app}

In this appendix we derive an alternative expression for the spectral overlap \eqref{MellinN_formula} that will be fundamental in computing the large-$N$ transseries expansion of the integrated correlator, $\corr$, starting from its spectral representation \eqref{C_spec_rep}. 
Throughout this derivation we assume for simplicity that $N\geq 2$ is an integer (which is also the case of physical interest) and that the spectral parameter $s$ lies on the critical line $\mbox{Re}(s)=\tfrac{1}{2}$. Despite these assumptions, the regime of validity for the final result \eqref{3F2_identity} will be more general and in particular it will provide an analytic continuation valid for $N\in \mathbb{C}$ with $\mbox{Re}(N)>0$.

\begin{figure}
    \centering
    \includegraphics[scale=0.17]{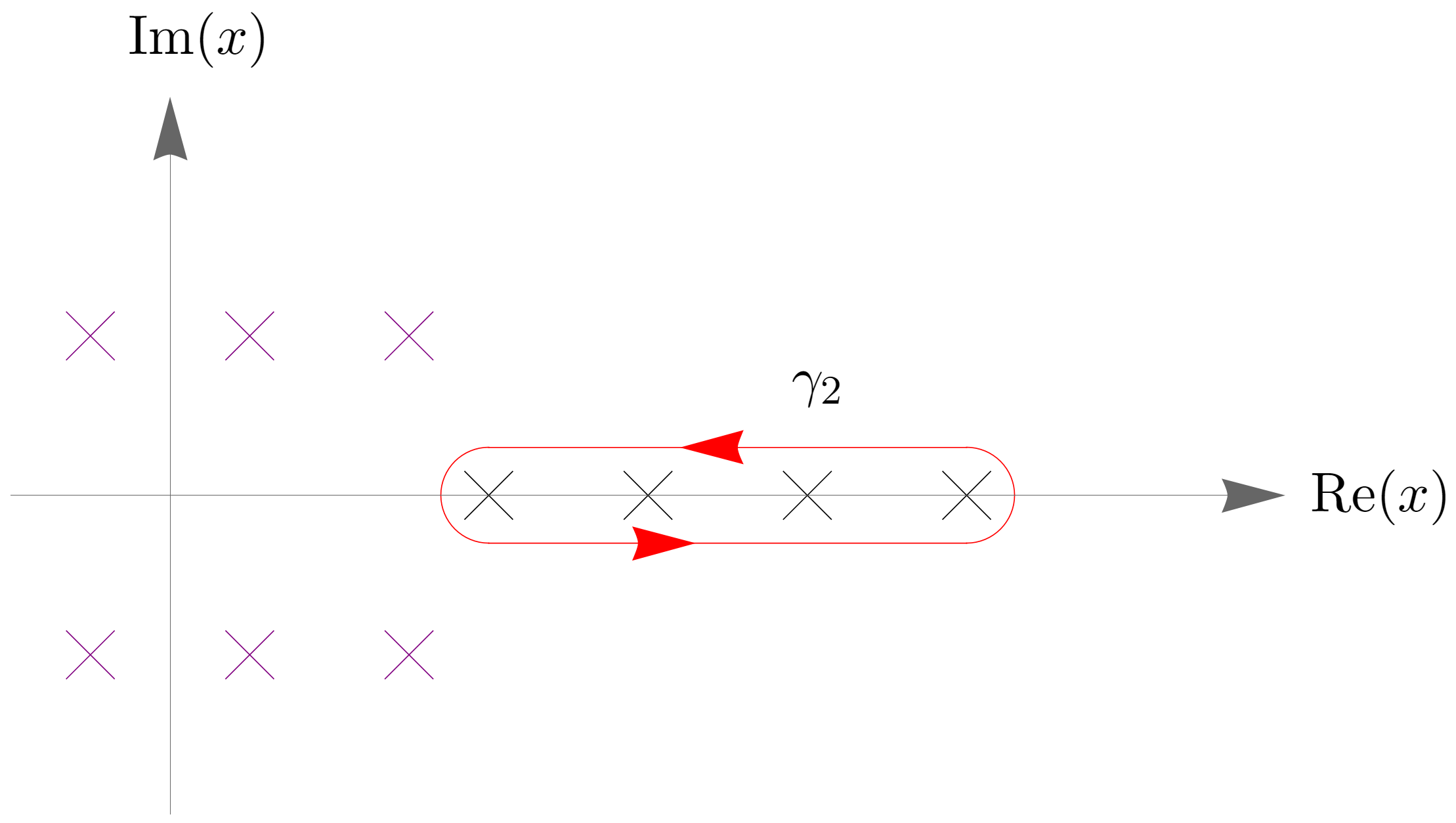}
    \caption{The contour of integration $\gamma_2$ circles around the poles in the complex $x$-plane located at $x\in\{2,3,...,N\}$ while avoiding other singularities.}
    \label{fig:gamma2}
\end{figure}

We begin by rewriting the hypergeometric function appearing in the spectral overlap \eqref{MellinN_formula} via the integral representation,
\begin{align}\label{first_cont_int_rep}
  &\notag  {}_3F_2(2-N,s,1-s;3,2\vert 1)\\
  &=\frac{2(-1)^N\Gamma(N-1)}{\Gamma(1-s)\Gamma(s)} \oint_{\gamma_1}\frac{\Gamma(x+N-s-1)\Gamma(x+N+s-2)\Gamma(x)}{\Gamma(x+N-1)\Gamma(x+N)\Gamma(x+N+1)}\frac{\dd x}{2\pi i}\,,
\end{align}
where $\gamma_1$ is a contour around the poles coming from $\Gamma(x)/\Gamma(x+N-1)$ and located at $x\in\{0,-1,...,-(N-2)\}$.  It is convenient to make a change of integration variables $x\to x'= x+N$ which we then rename $x$ again and, after using the reflection formula for the gamma functions, we reduce the integral to 
\begin{align}
  &\notag   {}_3F_2(2-N,s,1-s;3,2\vert 1)\\
  &\label{3F2_expr} =2\Gamma(N-1) \sin{(\pi s)} \oint_{\gamma_2}\frac{\Gamma(x-s-1)\Gamma(x+s-2)}{\sin{(\pi x)}\Gamma(N+1-x)\Gamma(x-1)\Gamma(x)\Gamma(x+1)}\frac{\dd x}{2\pi i}\,,
\end{align}
where $\gamma_2$ is a contour around the poles at $x\in\{2,3,...,N\}$ presented in figure \ref{fig:gamma2}. 

We now exploit the known asymptotic expansion of the gamma functions to find that the integrand of \eqref{3F2_expr} behaves as $|\mbox{Im}(x)|^{-3-N}$  as $|\mbox{Im}(x)|\to \infty$. Since this bound is uniform throughout the $x$-plane, we have that the integrals coming from the horizontal contributions to $\gamma_2$ located at $|\mbox{Im}(x)|=M$ vanish when we send $M\to \infty$.
The non-vanishing contributions to the contour integral \eqref{3F2_expr} can then be rewritten as
\begin{equation}
    \oint_{\gamma_2}=-\int_{\Re(x)=2-\epsilon}+\int_{\Re(x)=N+\epsilon}\,,
\end{equation}
with $\epsilon>0$ small enough.

 Furthermore, given that the integrand of \eqref{3F2_expr}  does not have any poles in the domain $\Re(x)>N+\epsilon$, we can push the second contour of integration towards~$\mbox{Re}(x) = +\infty$ and show that it vanishes given the bound on the integrand discussed above. We then deduce that the original contour integral~\eqref{3F2_expr} reduces simply to an integral over~$\Re(x)=2-\epsilon$,
 \begin{align}
  &\notag  {}_3F_2(2-N,s,1-s;3,2\vert 1)\\
  &\label{3F2_expr_final}=
-  2\Gamma(N-1) \sin{(\pi s)} \int_{\Re(x)=2-\epsilon} \frac{\Gamma(x-s-1)\Gamma(x+s-2)}{\sin{(\pi x)}\Gamma(N+1-x)\Gamma(x-1)\Gamma(x)\Gamma(x+1)}\frac{\dd x}{2\pi i}\,.
\end{align}

We now push the contour of integration towards $\mbox{Re}(x)\to -\infty$ and collect the residues from the poles originating from the two gamma functions at numerator, which are located at $x=-k+1+s$ and $x=-k+2-s$ for $k\in \mathbb{N}$, while the contribution at infinity vanishes thanks to a similar argument as above. Picking up these residues we arrive at the identity
\begin{align}\label{3F2_identity}
  & {}_3F_2(2-N,s,1-s;3,2\vert 1) = 2\Gamma(N-1) \times\\
  &\notag\Big[\frac{\Gamma(1{-}2s){}_3F_2(s{-}2,s{-}1,s;2s,N{+}s{-}1 \vert1)}{\Gamma(s+N-1)\Gamma(1-s)\Gamma(2-s)\Gamma(3-s)}{+}\frac{\Gamma(2s{-}1){}_3F_2({-}s{-}1,{-}s,1{-}s;2{-}2s,N{-}s \vert 1)}{\Gamma(N-s)\Gamma(s)\Gamma(s+1)\Gamma(s+2)}
    \Big]\,,
\end{align}
where we notice that the two factors in this expression are related by the transformation $s\to 1-s$, as expected given that the equation we started with \eqref{first_cont_int_rep} had this symmetry. 

Substituting \eqref{3F2_identity} back in the spectral overlap \eqref{MellinN_formula}, we derive the the symmetric expression 
\begin{align}\label{C_tilde_integral}
  &  M_N(s)= \\
  &\notag  \frac{2^{-2s}\sqrt{\pi}(2s-1)\Gamma(1+N)\Gamma(\frac{3}{2}-s)\tFt{s-2}{s-1}{s}{2s}{N+s-1}{1}}{\sin{(\pi s)}\Gamma(3-s)\Gamma(-s)\Gamma(N+s-1)} + (s\leftrightarrow 1-s)\,.
\end{align}
The above equation can be simplified even further by using the Euler-like integral representation
\begin{align}
    & \tFt{s{-}2}{s{-}1}{s}{2s}{N{+}s{-}1}{1}\label{3F2_integral_rep} {=}\frac{\Gamma(N{+}s{-}1)}{\Gamma(s)\Gamma(N{-}1)}   \int_0^1 (1{-}x)^{N-2} x^{s-1}\tFo{s{-}2}{s{-}1}{2s}{x}\dd x\,.
\end{align}
Combining \eqref{C_tilde_integral} with \eqref{3F2_integral_rep}, we conclude that 
\begin{align}
    M_N(s) \!&\label{eq:MNsApp}\phantom{:}= 
    \frac{2^{-2s} (2 s-1) \Gamma \left(\frac{3}{2}-s\right)}{\sqrt{\pi } \,\Gamma (-s)}
    \mathcal{I}_N(s)+(s\leftrightarrow 1-s) \,,\\ 
    \mathcal{I}_N(s) &\label{I_definition}\coloneqq \frac{N(N-1)}{(s-1)(s-2)}\int_0^1 (1-x)^{N-2}x^{s-1}\tFo{s-2}{s-1}{2s}{x}\dd x\,.
\end{align}
Finally, we rewrite the expression for $\mathcal{I}_N(s)$ by noticing that 
$$\frac{d^2}{dx^2}(1-x)^N = N(N-1)(1-x)^{N-2}\,,$$
which can be substituted in \eqref{I_definition} to perform twice an integration by parts.
However, we need to be careful in doing so since the boundary terms diverge as $x\to 0^+$  and we must introduce a regulator as to cancel the singular terms. The end result is a formula for the spectral overlap \eqref{eq:MNsApp} which is valid for $\Re(s)=\frac{1}{2}$ and $N\in \mathbb{C}$ with $\mbox{Re}(N)>0$
\begin{equation}\label{I_regularisation}
    \mathcal{I}_N(s) {=} \lim_{\epsilon\to 0^+}\Bigg[\frac{\epsilon^{s-2}}{s-2}+\frac{(s-2N-1)\epsilon^{s-1}}{2(s-1)}+\int_\epsilon^1 (1-x)^N x^{s-3}\tFo{s-1}{s}{2s}{x}\dd x\Bigg]\,.
\end{equation}

\section{Properties of a modular invariant Borel kernel}
\label{sec:AppE}

In this appendix we review some known properties for the modular invariant Borel kernel $\NPfn{t}$ as well as derive novel expressions which are of use in the main body of this work. 

The family of modular invariant functions $D_N(s;\tau)$ defined in \eqref{Dfn_def} was first introduced in \cite{Dorigoni:2022cua}, while a generalisation  has also recently appeared in the study of torodial Casimir energy in $3$-dimensional conformal field theories \cite{Luo:2022tqy}.
In this appendix we focus our attention to the special element \eqref{NPfn_def} in this family, namely  $\NPfn{t} =  D_{t^2}(0;\tau)$, which plays the r\^ole of a modular invariant kernel for our modified Borel resummation.

The Fourier mode decomposition of the lattice sum \eqref{NPfn_def} can be analysed straightforwardly \cite{Dorigoni:2022cua} starting from an integral representation valid for $\mbox{Re}(t^2)>0$ 
\begin{equation}
\NPfn{t}  = \sum_{(m,n)\neq(0,0)} \int_0^\infty e^{-x Y_{mn}(\tau)-\frac{4 t^2}{x}} \frac{2 t}{\sqrt{\pi } x^{3/2}} \dd x\,.\label{eq:EInt}
\end{equation}
In particular, for our analysis of the large-$N$ 't Hooft limit of section \ref{tHooft_section} we need an expression for the zero-mode sector of $\NPfn{t}$, which can be easily extracted from \eqref{eq:EInt} via standard Poisson resummation methods thus yielding
\begin{equation}
\int_{-\frac{1}{2}}^{\frac{1}{2}} \NPfn{t} \dd \tau_1 = \frac{2}{e^{4t\sqrt{\pi/\tau_2}}-1} + \sum_{n=1}^\infty 4n\tau_2 \,K_1(4n t\sqrt{\pi\tau_2})\,.\label{eq:Et0mode}
\end{equation}

Alternatively, the lattice sum representation \eqref{NPfn_def} can be written as a Poincar\'e series,
\begin{equation}
    \NPfn{t} =2\sum_{\gamma\in{\rm B}(\mathbb{Z})\backslash {\rm SL}(2,\mathbb{Z})} \mbox{Li}_0\Big(e^{-4t\sqrt{\pi/\tau_2}}\Big)= 2\sum_{\gamma\in{\rm B}(\mathbb{Z})\backslash {\rm SL}(2,\mathbb{Z})} \left[ \frac{1}{\exp{(4t\sqrt{\pi/\tau_2})}-1} \right]_\gamma\,.\label{eq:Poinc}
\end{equation}
Here $\gamma$ denotes an element of the coset space ${\rm B}(\mathbb{Z})\backslash {\rm SL}(2,\mathbb{Z})$, where ${\rm B}(\mathbb{Z})$ is the Borel stabilizer of the cusp $\tau = i\infty$, given by
\begin{equation}\label{Borel_defin}
    {\rm B}(\mathbb{Z}) \coloneqq \left\{ \pm\begin{pmatrix}  1 & n\\0 & 1\end{pmatrix} \,\middle|\, n\in\mathbb{Z}\right\} \subset {\rm SL}(2,\mathbb{Z})\coloneqq \left\{ \begin{pmatrix}  a & b\\c & d\end{pmatrix} \,\middle|\,a,b,c,d\in\mathbb{Z}\,,\,ad-bc=1\right\}\,.
\end{equation}
The notation $[f(\tau)]_\gamma \coloneqq f(\gamma\cdot \tau)$ in \eqref{eq:Poinc} indicates that the coset element $\gamma \in {\rm B}(\mathbb{Z})\backslash {\rm SL}(2,\mathbb{Z})$ acts  on any instance of the modular parameter $\tau $ via the standard fractional transformation 
\begin{equation}
\gamma\cdot\tau \coloneqq \frac{a\tau+b}{c\tau+d} \,,\qquad \gamma = \left( \begin{matrix}a & b \\ c & d\end{matrix}\right)\,.
\end{equation}

As shown in \cite{Paul:2023rka}, the  spectral representation of $\NPfn{t}$ can be obtained directly from the Poincar\'e series representation \eqref{eq:Poinc}
and takes the form 
\begin{equation}\label{NPfn_spect_rep}
    \NPfn{t} = \frac{1}{8t^2}+4\int_{\Re(s)=\frac{1}{2}} \frac{(4t)^{-2s}\Gamma(2s)}{\Gamma(s)}\eisen{s}\frac{\dd s}{2\pi i}\,.
\end{equation}

From the spectral representation we may also obtain a different expression for the Fourier zero-mode \eqref{eq:Et0mode}. We start by substituting in \eqref{NPfn_spect_rep} the Fourier zero-mode \eqref{Eisen_integral_rep} of the non-holomorphic Eisenstein series, 
\begin{equation}
\int_{-\frac{1}{2}}^{\frac{1}{2}} \eisen{s}  \dd \tau_1 = \xi(2s)\tau_2^s + \xi(2s-1)\tau_2^{1-s}\,.
\end{equation}
To evaluate \eqref{NPfn_spect_rep} we focus separately on the contribution coming from either of the two terms $\xi(2s)\tau_2^s$ and $\xi(2s-1)\tau_2^{1-s}$ in the above expression. 

Starting with the term  $\xi(2s)\tau_2^s$, we rewrite the completed Riemann zeta using the Dirichlet series representation of the zeta function $\zeta(2s)=\sum_{k\geq 1}k^{-2s}$ valid for $\mbox{Re}(s)>1/2$ 
\begin{align}
&\notag   4\int_{\Re(s)=\frac{1}{2}+\epsilon} \frac{(4t)^{-2s}\Gamma(2s)}{\Gamma(s)}\xi(2s) \tau_2^{s} \frac{\dd s}{2\pi i} 
  =\sum_{k=1}^\infty  4\int_{\Re(s)=\frac{1}{2}+\epsilon} \Gamma(2s) \Big(\frac{4 t \sqrt{\pi } }{\sqrt{\tau_2}}k\Big)^{-2s} \frac{\dd s}{2\pi i}\\
  &= \frac{2}{e^{4t\sqrt{\pi/\tau_2}}-1}\,,
\end{align}
where we simply evaluate the sum over all the residues coming from the poles of $\Gamma(2s)$ and subsequently perform the sum over $k$.
Note that in doing so we have to shift the contour of integration to the right by an infinitesimal quantity, $\epsilon>0$, since $\xi(2s)$ has a pole at $s=1/2$.
The contribution from this pole is cancelled by an equal and opposite pole coming from the second zero-mode term $\xi(2s-1)\tau_2^{1-s}$, since the complete non-holomorphic Eisenstein $\eisen{s}$ is perfectly regular for $s=\frac{1}{2}$ and the integral expression \eqref{NPfn_spect_rep} is unchanged if we move the contour of integration to $\mbox{Re}(s)=1/2+\epsilon$. 

The remaining zero-mode contribution to $\NPfn{t}$ comes from the leftover $1/(8t^2)$ term in \eqref{NPfn_spect_rep} and the non-holomorphic Eisenstein series factor $\xi(2s-1) \tau_2^{1-s}$ which can be combined as
\begin{align}
    \mathcal{U}(t;\tau_2)&\notag \coloneqq \frac{1}{8t^2}+\int_{\Re(s)=\frac{1}{2}+\epsilon} \frac{4(4t)^{-2s}\Gamma(2s)}{\Gamma(s)}\xi(2s-1)\tau_2^{1-s}\frac{\dd s}{2\pi i}\\
        &\label{U_definition} \,= \int_{\Re(s)=1+\epsilon} \frac{ \Gamma \left(s-\frac{1}{2}\right) \Gamma \left(s+\frac{1}{2}\right) \zeta (2 s-1)}{2\pi t^2} \left(2 t \sqrt{\pi  \tau_2}\right)^{2-2 s} \frac{\dd s}{2\pi i}\,.
\end{align}

We then conclude that the Fourier zero-mode of $ \NPfn{t}$ can also be expressed as
\begin{equation} \label{E0_formula}
   \int_{-\frac{1}{2}}^{\frac{1}{2}} \NPfn{t} \,\dd \tau_1 = \frac{2}{e^{4t\sqrt{\pi/\tau_2}}-1} + \mathcal{U}(t;\tau_2)\,,
\end{equation}
where it is worth noting that for large values of $t$ both terms are exponentially suppressed.
Comparing with \eqref{eq:Et0mode}, we see that $\mathcal{U}(t;\tau_2)$ has the alternative representation
\begin{equation}
\mathcal{U}(t;\tau_2) = \sum_{n=1}^\infty 4n\tau_2 \,K_1(4n t\sqrt{\pi\tau_2})\,.\label{eq:UBessel}
\end{equation} 
Unfortunately $\mathcal{U}(t;\tau_2)$  does not seem to have a simpler expression in terms of elementary functions, however we can use the integral representation for the Bessel function to rewrite \eqref{eq:UBessel} in the useful Borel-like form
\begin{align}
\mathcal{U}(t;\tau_2) 
= \sum_{n=1}^\infty  \int_t^\infty (4n \tau_2) e^{- 4 n x \sqrt{\pi \tau_2}} \frac{x}{ t \sqrt{x^2-t^2}}\, \dd x\,.\label{eq:UBorel}
\end{align}

To discuss the 't Hooft large-$N$ limit in section \ref{tHooft_section}, we need certain integrals involving the Fourier zero-mode just discussed.
In particular, we need a formula for the moments $t^\alpha$ with respect to the measure $\mathcal{U}(t;\tau_2) \dd t$.
From the definition \eqref{U_definition}, we see that the $t$-dependence of such integrals is simply of the form $t^{\alpha-2s}$, we then consider the analytic continuations
\begin{align}
    \int_0^1 t^{\alpha-2s} \, \dd t &=\frac{1}{\alpha-2s+1}\,, \qquad\,\,\,\,\,  \quad \Re(s)<\frac{\alpha+1}{2}\,,\\
   \int_1^\infty t^{\alpha-2s}\,\dd t&=-\frac{1}{\alpha-2s+1}\,, \qquad \quad \Re(s)>\frac{\alpha+1}{2}\,.
\end{align}
After having performed the integral over $t$, we notice that the integrand of \eqref{U_definition} acquires a single simple pole located at $s=\frac{\alpha+1}{2}$. Closing the contour of integration to the right half-plane $\mbox{Re}(s)>1$ picks up the residue at this pole without any boundary contribution, thus giving us
\begin{equation}\label{U_integral_formula}
    \int_0^\infty  \mathcal{U}(t;\tau_2)\, t^{\alpha} \,\dd t = \frac{ \Gamma\Big(\frac{\alpha}{2}+1\Big)\Gamma\Big(\frac{\alpha}{2}\Big)\zeta(\alpha)}{4\pi} (2\sqrt{\pi \tau_2})^{1-\alpha}\,.
\end{equation}

Alternatively, we may start from the expression in terms of Bessel functions \eqref{eq:UBessel} and compute
\begin{equation}
\int_0^\infty 4 n \tau_2 K_1(4 n t \,\sqrt{\pi \tau_2}) \,t^\alpha\,\dd t = \frac{\Gamma \Big(\frac{\alpha }{2}+1\Big) \Gamma \Big(\frac{\alpha }{2}\Big) }{4 \pi  } \,n^{{-\alpha}}\,(2 \sqrt{\pi  \tau_2})^{1-\alpha }\,,
\end{equation} 
valid for $\mbox{Re}(\alpha)>0$, which can then be easily summed over $n$ to reproduce \eqref{U_integral_formula}.

In section  \ref{tHooft_section} we also discuss the non-perturbative terms in the dual 't Hooft limit which require finding an expression for the moments $t^\alpha$ with respect to the measure $\mathcal{U}(t+1;\tau_2) \dd t$. To this end, we start with the identity
\begin{equation}
    \int_0^\infty \frac{t^\alpha}{(t+1)^{2s}}\, \dd t=\frac{\Gamma(1+\alpha)\Gamma(2s-1-\alpha)}{\Gamma(2s)}\,,
\end{equation}
which we then substitute in the defining formula \eqref{U_definition} to derive
\begin{equation}
    \int_0^\infty \mathcal{U}(t+1;\tau_2)\, t^{\alpha}\,\dd t = \Gamma(1+\alpha) \int_{\Re(s)=\frac{1+\alpha}{2}+\epsilon}\frac{\Gamma(s-\frac{1}{2})\Gamma(2s-1-\alpha)\zeta(2s-1)}{4\sqrt{\pi}\Gamma(s)(4\sqrt{\pi\tau_2})^{2s-2}}\frac{\dd s}{2\pi i}\,.
\end{equation}
Although this expression does not quite yield a closed form such that \eqref{U_integral_formula}, it still suffices for the discussion of section  \ref{tHooft_section}.

\bibliographystyle{utphys}
\bibliography{cites}

\end{document}